\input psfig
\documentstyle
[referee]   
{l-aa}	    
\begin{document}
\def\lya{Ly$\alpha$}     %
\def\lyb{Ly$\beta$}      %
\def\Wr{w_{\rm r}}       %
\def\mr{m_{\rm r}}       %
\def\Mr{M_{\rm r}}       %
\def\kms{~km~s$^{-1}$}   %
\def\za{z_{\rm a}}       %
\def\zg{z_{\rm g}}       %
\def\h50{h_{50}^{-1}}    %
\thesaurus {03(11.17.1; 11.05.2; 11.09.3)}
\title{Observational constraints on the nature of low redshift \lya
\ absorbers\thanks{Based on observations made at the
Canada-France-Hawaii-Telescope, Hawaii, USA}}
\author{V. Le Brun \inst{1} \and J. Bergeron \inst{2,1} \and P. Boiss\'e
\inst{3}}
\institute {Institut d'Astrophysique de Paris, CNRS, 98bis boulevard Arago,
F-75014 Paris, France
\and European Southern Observatory, Karl-Schwarzschild Stra\ss e 2, D-85748
Garching, Germany
\and Ecole Normale Sup\'erieure, 24 rue Lhomond, F-75005 Paris, France}
\offprints {V. Le Brun}
\date      {Received 13 January, accepted 6 June 1995}
\maketitle
\markboth{V. Le Brun et al.: Constraints on the nature of the \lya \
absorbers}{}
\begin{abstract}
We present results from a spectroscopic and imaging survey
of galaxies in the fields of quasars from the Hubble Space Telescope (HST)
Quasar Absorption Line Key Project. The aim of this survey is to identify
galaxies  within 3.5\arcmin \ from the quasar sightline, to a limiting,
integrated $r$-band magnitude $\mr = 22.5$. The data are then
compared  to the HST homogeneous sample of \lya-only absorbers in order to
put constraints on the nature of these absorbers, and in particular on
their relation to galaxies.\par
We have obtained spectra for 81 objects in three quasar fields and identified
66 galaxies (success rate of 81\%) at redshifts in the range $z=0.0500$-0.7974,
and at linear impact parameters to the quasar sightlines spanning $D$=57-2380~
$\h50$ kpc. Among these galaxies, 19 are at less than 750\kms \ from a \lya \
absorber, and only one clearly does not give rise to any absorption. Three
other
galaxies are at the redshifts of metal-rich absorption systems, of which one
belongs to a cluster with altogether 19 identified galaxies at the quasar
redshift.\par
The analysis of our sample combined with those of previous studies shows that:
\par\noindent
1) the redshift agreements of the \lya \ absorber-galaxy associations
cannot be due to chance coincidence;\par\noindent
2) there is no clear anti-correlation between
the \lya \ rest-frame equivalent width and the impact parameter for the whole
sample ($w_{\rm r, min}$=0.10 \AA). When only the strongest lines are
considered
($\Wr \ge 0.24$~\AA), $\Wr \left({\rm Ly}\alpha \right)$ and $D$ are
marginally
anti-correlated.
Lanzetta et al. (1995) found a stronger anti-correlation which could be due, at
least in part, to the existence of metal-rich absorbers in their sample. Our
results suggest that most \lya \ absorbers are not gaseous clouds that belong
in
a strict sense to galaxies, as is the case for Mg\,{\sc ii} absorbers. The size
of \lya \ galactic halos can be inferred from the variation with $D$ of the
fraction of associations to the total number of galaxies at impact parameters
$< D$. This fraction drops from 1 to $\sim$ 0.65 at $ D \sim 200 \h50$ kpc and
flattens at larger values of $D$ ($\ga 300 \h50$ kpc).
This leads to \lya \ galactic halos sizes about
three times larger than the  inner Mg\,{\sc ii} halo region;\par\noindent
3) there is no correlation between the galaxy luminosity and the impact
parameter. This again suggests that a significant \lya \ clouds do not belong
to individual galaxies, but instead are distributed in the local large-scale
structure. For the smaller impact parameters, this could reflect a link between
$D$ and the total galaxy mass rather than its luminous mass;\par\noindent
4) the HWHM=120 \kms \ of the relative velocity distribution of the \lya \
absorber-galaxy associations is consistent with either galaxy rotation
velocities or the local velocity dispersion in large-scale
structures.
\keywords{quasars: absorption lines -- Galaxies: evolution -- intergalactic
medium}
\end{abstract}
\section{Introduction}
The homogeneous samples of \lya \ and C\,{\sc iv} absorption lines
detected in quasar spectra observed with the Hubble Space Telescope (HST)
are of primary importance to investigate the properties of the
intergalactic medium and the gaseous content of galaxies at low redshift.
These UV spectra when combined with
ground-based observations of higher redshift absorption lines, provide a data
base which allows to determine the evolution in cosmic time of gaseous
objects covering a very large range in H\,{\sc i} column density.
\par
Analysis of the results from the HST absorption line survey (Bahcall et al.
1991, 1993, 1994) shows that the  ``\lya \ absorbers'' (with absorption lines
only from the Lyman series) constitute the dominant population at all
redshifts and that their number density at low redshift is somewhat higher than
expected from an extrapolation of the data obtained in the range
$1.7<z<3.5$ (Bahcall et al. 1993, 1994). Groups of \lya \ lines
are found  close in redshift to about half of the extensive metal-line systems
(Bahcall et al. 1994).
\par
 From a study of the \lya \ absorber-galaxy correlation
function, Morris et al. (1993) have suggested that all galaxies
more luminous than 0.1 $L^\star$ have effective cross-sections of 0.5-1~
$\h50$~Mpc, where $h_{50}$ is the Hubble constant in units of
50\kms ~Mpc$^{-1}$ and the deceleration parameter is assumed to be equal to
zero. This however does not imply a physical association between the \lya \
absorbers and individual galaxies. Lanzetta et al. (1995, hereafter LBTW) have
found an anti-correlation between the impact parameter of the absorption-line
selected galaxies and the rest equivalent width of the \lya \ absorption line
which suggests a tight physical link between absorbers and galaxies. These
authors conclude that most luminous galaxies are surrounded by extended
galactic halos of radius $r \sim 345 \h50$~kpc (using q$_0$ = 0 and $z$ = 0.3).
Recently, Morris \& van den Bergh (1994) have proposed that
tidal effects may play an important role in extracting gas from
the disc of galaxies and spreading it over hundreds of kiloparsecs. It is
indeed likely that at least some of the \lya \ systems are caused by tidal
debris which remain loosely associated with the
parent galaxies (assuming that the gas is leaving the galaxy with a velocity
of 100\kms \ in excess of the escape velocity, the expected linear
separation after 3 $\times$ 10$^9$ yr is about $300 \h50$~kpc). In this
picture, the intervening gas is supposed to be ``physically associated''
to the galaxy, in the sense that it originated from it. However,
for large impact parameters, the link between \lya \ absorbers and galaxies
could be only apparent if  the \lya \ absorptions were due to gaseous clouds
distributed in space in the same way as dark matter (Cen et al. 1994;
Petitjean et al. 1995). Such a situation could result from a partial
inefficiency in  the galaxy formation process that would have left clumps of
metal-poor gas displaying the same large-scale distribution as galaxies.
\par
As at high redshift, the C\,{\sc iv} systems are the dominant population of
metal-rich absorbers at $\za \sim 0.5$ (Bahcall et al. 1993)
and often show associated O\,{\sc vi} absorption (Lu \& Savage 1993; Bahcall
et al. 1993, 1994).
Over the redshift range $z \simeq$ 0.15-1.05, Mg\,{\sc ii} absorption systems
have been identified  with large gaseous halos around bright
($-24 < M_{\rm r} < -20$) galaxies (Bergeron \& Boiss\'e 1991 and
references therein; Bergeron et al. 1992; Steidel 1993). There is a
correlation between the radius of these Mg\,{\sc ii} halos and the galaxy
luminosity, $R \simeq 75 (L/L^\star)^\alpha \h50$ kpc, where $\alpha \simeq$
0.2 to 0.3  (Bergeron \& Boiss\'e 1991; Le Brun et al. 1993; Steidel 1993).
The linear scales probed by Mg\,{\sc ii}
absorbers is thus much smaller than those typical for \lya \ absorbers.
Few C\,{\sc iv} absorption systems  have been so far identified, but they
most probably arise in extended galactic halos of size similar to those of
Mg\,{\sc ii} absorbers or possibly slightly larger. One of the
Mg\,{\sc ii}-absorption selected galaxy has associated  C\,{\sc iv} and
O\,{\sc vi} absorptions  (Bergeron et al. 1994), and two other C\,{\sc iv}
systems have recently been identified by LBTW. The impact parameters of these
three C\,{\sc iv}-absorption selected galaxies  cover the range
$d =$ (58-95)$\h50$ kpc. The high ionization O\,{\sc vi} absorbers could be an
homogeneous phase of low density and  large extent in which  C\,{\sc iv} and
Mg\,{\sc ii} clouds are embedded. Galactic halos around bright galaxies are a
common phenomenon at moderate redshift and their most outer parts might only be
probed by \lya \ absorbers.
\par
The identification of a large sample of low redshift \lya \ systems
is needed to investigate in more details the nature of these absorbers and
determine which fraction of them correspond to gas that is truly associated
with
galaxies. To this aim, we have undertaken a spectroscopic survey of galaxies
around quasars of the HST Key Project. The  study of three fields, reported in
this paper, substantially increases the
size of the sample of galaxies in the environment of \lya \ absorbers.
We have searched for possible correlations involving the impact parameter of
the galaxy, its luminosity, the \lya \ absorber-galaxy velocity difference
and the equivalent width of the \lya \ absorption line in order to constrain
the nature of \lya \ absorbers: gravitationally bound very extended galactic
halos, gas extracted by tidal interaction or intergalactic gas that traces
the distribution of luminous and dark matter in the universe.
 \par
The observations and data reduction are presented in Sect.~2. The list of
identified galaxies and the description of
individual fields are given in Sect.~3. Results of  statistical analyses
of the combined sample, which includes studies from other groups, are presented
in Sect.~4. In Sect.~5, we discuss the implications of these results on
the nature of the \lya \ absorbers and present our conclusions.
\section{Observations and data reduction}
The log of the observations is given in Table~\ref{obslog}.
\begin{table*}
\caption{Journal of the observations}
\label{obslog}
\begin{tabular}{llllll}
\hline\noalign{\smallskip}

Object    & Dates              & Instrument & mode         & Filter   &
n $\times \Delta t$ \\
          &                    &            &              & or Grism &
        mn             \\
Ton 153   & 1992 March, 27     & FOCAM      & Imaging      &   r      &
  $3 \times 7$         \\
          & 1993 May 18,20,21 & MOS-SIS     & Imaging      &   r      &
  $5 \times 5$         \\
          & 1993 May 20,21    & MOS-SIS     & Multi Spect. & O 300    &
  $3 \times 90$        \\
3C~351    & 1992 March, 27    & FOCAM       & Imaging      &    r     &
  $4 \times 5$         \\
          & 1993 May 17,19    & MOS-SIS     & Imaging      &    r     &
  $3 \times 5^{\rm a}$ \\
          & 1993 May 19       & MOS-SIS     & Multi Spect. & O 300    &
  $2 \times 90$        \\
H~1821+6419 & 1992 Sept 28      & HRCAM       & Imaging      &    r     &
  $3 \times 10$        \\
          & 1993 May 16,17,18 & MOS-SIS     & Imaging      &    r     &
  $3 \times 5^{\rm a} $\\
          & 1993 May 17,18    & MOS-SIS     & Multi Spect. & O 300    &
  $3 \times 90$  \\
\noalign{\medskip}\hline
\end{tabular}
\smallskip\noindent \\
$^{\rm a}$ Only two exposures were used to obtain the final image due to a
displacement of the CCD on the 18$^{th}$.
\end{table*}
\subsection{Imaging}
The images were obtained at the prime focus of the CFH Telescope with the
 Faint Object Camera (FOCAM) or the High Resolution Camera (HRCAM). The CCD is
a
640$\times$1024 pixel RCA with a plate scale of 0.2055\arcsec \  per pixel for
FOCAM and a 1080$\times$1080 pixel SAIC with a plate scale of 0.1311\arcsec \
per pixel for HRCAM. The ESO
software package MIDAS was used at all steps of the data processing. The images
were reduced in a standard way, with subtraction of individual overscan and
of a median filtered bias before flat-fielding. The flat-field was obtained by
an averaging procedure including pixel rejection, applied to all the images
(at least 16) with a similar background level. The resulting image was then
normalized. A shift-and-add procedure was used, and the averaging obtained
with a median filter allowed a successful removal of  cosmic ray events.
\par
Other images were taken with the CFHT Multi-Object Spectrograph (MOS) in the
imaging mode. The CCD is a (2048)$^2$ pixel Lick~3 with a plate scale
of 0.3140\arcsec \  per pixel. The actual  field imaged is 9.7\arcmin
$\times$ 9.4\arcmin. The same reduction procedures have been applied; in
addition, a distortion map was used to correct for the
optical distortion which is fairly strong in the field corners. Due to this
optical distortion and the poorer sampling of MOS, the  images obtained with
this instrument are only used  to study the galaxies further away to the
quasar line of sight, the
closest ones being better investigated with the higher-quality FOCAM/HRCAM
images.
\begin{table*}
\caption{Galaxy population of the fields}
\label{galpop}
\begin{tabular}{rrrrrrcrrrcr}
\hline\noalign{\smallskip}
                & \multicolumn{3}{c}{TON~153}  &\ & \multicolumn{3}{c}{3C~351}&
\
                & \multicolumn{3}{c}{H~1821+6419} \\
\noalign{\smallskip}
\cline{2-4}\cline{6-8}\cline{10-12}
\noalign{\smallskip}
Magnitude range & N$_{\rm gal}^{\rm a}$ & N$_{\rm z}^{\rm b}  $ & Comp.$^{\rm
c}$ & &
	          N$_{\rm gal}^{\rm a}$ & N$_{\rm z}^{\rm b,d}$ & Comp.$^{\rm c}$ & &
		  N$_{\rm gal}^{\rm a}$ & N$_{\rm z}^{\rm b,e}$ & Comp.$^{\rm c}$ \\
\noalign{\smallskip\hrule\medskip}

$ \ \ \ \mr \le 17.5$&  1 & 0 &  0\% & &  4 & 1 (+2) & 75\% & &  1 & 1 (+0)
&100\% \\
$17.5 < \mr \le 18.5$&  2 & 2 & 67\% & &  4 & 2 (+2) & 87\% & &  6 & 1 (+1) &
43\% \\
$18.5 < \mr \le 19.5$&  1 & 0 & 40\% & &  4 & 2 (+0) & 75\% & & 30 & 7 (+5) &
40\% \\
$19.5 < \mr \le 20.5$&  9 & 2 & 31\% & &  9 & 1 (+1) & 52\% & & 67 & 6 (+1) &
21\% \\
$20.5 < \mr \le 21.5$& 25 & 9 & 34\% & & 54 & 4 (+1) & 21\% & &124 & 3 (+1) &
11\% \\
$21.5 < \mr \le 22.5$& 58 & 5 & 19\% & &120 & 5 (+0) & 11\% & &158 & 0 (+0) &
7\% \\
\noalign{\smallskip}\hline
\end{tabular}
\\ \smallskip\noindent
$^{\rm a}$ N$_{\rm gal}$ is the number of galaxies in each magnitude bin within
3.5\arcmin \ of the quasar sightline   \noindent \\
$^{\rm b}$ N$_{\rm z}$ is the number of galaxies in each magnitude bin with
measured redshift   \noindent \\
$^{\rm c}$ Completeness, it refers to the cumulated distribution of magnitudes
\noindent \\
$^{\rm d}$ The additional numbers in brackets show the redshifts obtained by
LBTW \noindent \\
$^{\rm e}$ The additional numbers in brackets show the redshifts obtained by
Schneider et al. (1992)
\end{table*}
\subsection{Spectroscopy}
Galaxy spectra have been obtained with MOS in the multi-slit spectroscopy
mode. We have observed between 25 and 28 galaxies per mask with slitlets of
15 to 18\arcsec \ length and 1.5\arcsec \ width. After correction of the
distortion, overscan and bias subtraction, the spectra have been divided by an
averaged dome flat-field.
The  grating  used gives a useful, maximum  wavelength range  of
$4400 \lid \lambda \lid 9700$~\AA, a wavelength dispersion of 3.6~\AA \
per pixel and a resolution of FWHM = 18~\AA \ for a 1.5\arcsec \  slit-width.
The wavelength calibration was done using Helium and Argon spectra.
Spectra of standard stars were taken in one of the slitlets of different masks
for flux calibration. Thereafter, we have developed  semi-automatic procedures
to reduce the multiple spectra. For the wavelength calibration, identification
of only one set of He or Ar lines is needed for a given mask, e.g. in the
first slitlet, the recognition being then  made automatically in each of the
other slitlets. For individual exposures of a given field, the sky is fitted
along the spatial rows by a polynomial of low order, after a strong filtering
to avoid spurious features due to cosmic ray events. The extraction of the
spectra is done with a weighted average of the spatial rows. The cosmic ray
hits, present on the galaxy spectra themselves, are removed by comparing the
spectra obtained from different exposures.
\par
The redshifts have been measured using strong emission and absorption lines,
when present. For the other galaxies, the cross-correlation task ``xcsao'' of
IRAF has been used. We have determined 66 definite galaxy redshifts for the
three fields, thus a success rate of 81\%.
\section{Identification of the galaxies and description of the individual
fields}
To search for galaxies (possibly) related to \lya \ absorbers, we first need to
define a criterion involving the relative velocity  between the absorbers and
the galaxies, $\Delta v$. Let us first discuss the maximum absolute value of
$\Delta v$ which would be appropriate to associate gas to a given galaxy. For
ellipticals, the average velocity dispersion is about 200\kms, and can reach
300\kms; the rotational velocity is smaller and its maximum value is around
150\kms \ (de Zeeuw \& Franx 1991). The radial velocity in spirals is
correlated
to the galaxy luminosity. In normal galaxies, the velocity at large radii is of
the order of 200-250\kms, and can reach 350\kms \ for the most luminous
galaxies
(Rubin et al. 1982). In starburst-driven galactic superwinds, the total
velocity
variation along extended emission-line region is of order 600\kms, and in some
cases is as large as 900\kms \ (Heckman et al. 1990). Within large-scale
structures, the typical velocity spread between galaxies within a sheet-like
structure is about 500\kms, and may reach 750\kms \  (Ramella et al. 1989).
Given
the uncertainties in the redshift measurements of both the \lya \ absorption
lines
and the field galaxies, $| \Delta z |\le 0.0003$ for our sample as well as for
that of Morris et al. (1993) and LBTW, we have adopted as maximum, relative
velocity between the \lya \ absorber and the ``associated galaxy'' $|\Delta v|$
=
750\kms.
\par
For comparison, the largest values of  $|\Delta v|$  in the sample
of Morris et al. (1993) is 710\kms \ and the relative velocity criterion
adopted by LBTW is  $|\Delta v| <$ 1000\kms \  to
include absorber-galaxy associations on scales up to those of clusters of
galaxies. In  the LBTW GA sample, restricted to unambiguous
redshift identifications,  the largest value of  $|\Delta v|$ is just below
500\kms.
\par
The total number of galaxies within 3.5\arcmin \ of a quasar sightline in
the different magnitude bins, the number of galaxies with measured redshifts
and
the completeness for the cumulative galaxy distribution are given in
Table~\ref{galpop} for the three quasar fields. The redshifts, magnitudes and
distances ($\Delta$x=$-\Delta\alpha$, $\Delta$y=$\Delta\delta$)
to the quasar sightlines  of the galaxies with spectroscopic data are
listed in Tables~\ref{Tontab}, \ref{3Ctab} \ and \ref{H18tab} for
the fields around TON~153, 3C~351 and H~1821+6419, respectively. The \lya \
system(s) within 750\kms \ of the galaxy redshift are listed in Column 7, and
the
comments given in Column 8 refer to the nature of the absorption system,
atmospheric features in the quasar spectra occuring at the expected wavelength
of the \lya \ absorption possibly associated with the galaxy or the lack of
available data at the  galaxy redshift.

\begin{table*}
\caption{Characteristics of the TON~153 field}
\label{Tontab}
\begin{tabular}{lrrrrlllll}
\hline\noalign{\smallskip} \\
Gal &$\Delta$x(\arcsec )&$\Delta$y(\arcsec )&$\theta$(\arcsec )&d(kpc)&$\zg$&
$\za$ & Observations &
 $\mr$   & $\Mr$ \\
\hline \\
QSO&    0.0 &    0.0 &   0.0 &    0.0 & 1.0220 &                &
 & 16.0 & -28.9 \\
1  &  267.4 &   19.5 & 268.1 &  920.3 & 0.1442 &                & unobs.$^{\rm
a}$
 & 19.9 & -19.9 \\
2  &  255.3 &  -28.4 & 256.8 & 1263.1 & 0.2291 &                & unobs.$^{\rm
a}$
 & 19.3 & -21.7 \\
3  &  232.0 &   25.4 & 233.4 & 2013.7 & 0.5674 &                & Coinc.$^{\rm
b}$
 & 21.5 & -21.7 \\
4  &  208.9 & -118.7 & 240.3 & 2375.7 & 0.7663 & 0.7667         & \lya \
 & 22.7 & -21.3 \\
5  &  192.2 &   29.3 & 194.4 & 1624.0 & 0.5324 & 0.5349         & \lya \
 & 21.7 & -21.3 \\
6  &  176.0 &   -3.3 & 176.0 & 1669.7 & 0.6952 & 0.6947         & \lya \
(Bl.$^{\rm c}$)
 & 21.8 & -22.0 \\
7  &  163.8 &   23.5 & 165.5 &  223.9 & 0.0500 &                & unobs.$^{\rm
a}$
 & 18.4 & -19.0 \\
8  &  142.5 &  -91.1 & 169.1 &  423.3 & 0.0992 &                & unobs.$^{\rm
a}$
 & 21.0 & -18.0 \\
9  &  118.7 &  -26.1 & 121.5 &  770.1 & 0.3310 &                & unobs.$^{\rm
a}$
 & 19.7 & -22.1 \\
10 &   99.7 &  -41.4 & 107.9 &  766.2 & 0.3976 &                & Coinc.$^{\rm
d}$
 & 22.2 & -20.1 \\
12 &   78.9 &  -69.2 & 105.0 &  996.9 & 0.6974 & 0.6958         & \lya \
(Bl.$^{\rm c}$)
 & 22.3 & -21.4 \\
13 &   56.9 &  -78.3 &  96.8 &  634.5 & 0.349  &                & unobs.$^{\rm
a}$
 & 20.6 & -21.3 \\
14 &   42.4 &  -62.6 &  75.6 &  438.1 & 0.2891 & 0.2891         & metal
 & 20.4 & -21.1 \\
15 &   19.8 & -108.4 & 110.1 &  472.6 & 0.1910 &                & unobs.$^{\rm
a}$
 & 18.0 & -22.5 \\
16 &    6.8 &    4.8 &   8.3 &   77.5 & 0.6715 & 0.6716, 0.6736 & \lya \
 & 21.7 & -22.0 \\
17 &  -10.9 &   -7.8 &  13.4 &        & indef. &                &
 & 23.0 &       \\
18 &  -25.6 &  -96.4 &  99.7 &        & 0.0000 &                &  star
 & 21.3 &       \\
19 &  -44.4 &   -2.3 &  44.5 &  415.3 & 0.6717 & 0.6716, 0.6736 & \lya \
 & 22.3 & -21.3 \\
20a&  -66.1 &   10.5 &  67.0 &  563.5 & 0.5397 & 0.5436         & \lya \
(Bl.$^{\rm e}$)
 & 20.8 & -22.3 \\
20b&  -68.1 &    9.7 &  68.7 &  578.4 & 0.5398 & 0.5436         & \lya \
(Bl.$^{\rm e}$)
 & 21.1 & -22.0 \\
21 &  -82.3 &   32.6 &  88.5 &        & indef. &                &
 & 21.3 &       \\
22 &  -96.9 &   25.7 & 100.3 &        & indef. &                &
 & 22.8 &       \\
23 & -114.5 & -107.4 & 157.0 & 1421.1 & 0.6265 & 0.6267         & \lya \
(Bl.$^{\rm f}$)
& 20.7 & -22.8 \\
24 & -129.5 &  -31.5 & 133.3 & 1242.4 & 0.668  & 0.6692         & \lya \
 & 20.5 & -23.2 \\
25 & -155.6 &  -29.1 & 158.3 & 1471.1 & 0.6639 & 0.6606         & metal
 & 21.3 & -22.4 \\
26 & -170.2 &    6.6 & 170.3 &  739.9 & 0.1944 &                & unobs.$^{\rm
a}$
 & 20.5 & -20.0 \\
27 & -194.4 &   13.4 & 194.8 &        & indef. &                &
 & 20.7 &       \\
28 & -213.5 &  -23.3 & 214.8 &        & indef. &                &
 & 22.4 &       \\
\noalign{\smallskip}\hline
\end{tabular}
\smallskip\noindent \\
$^{\rm a}$ At these redshifts, there are no available HST data on \lya \
absorption.\\ \noindent
$^{\rm b}$ The  \lyb \ line at 1902.02~\AA \ is at less than 750\kms \ from
the galaxy redshift (see text).
\\ \noindent
$^{\rm c}$ These two absorption redshifts come from a two-component fit to the
\lya \ line at 2060.88~\AA \ (see text).
\\ \noindent
$^{\rm d}$ The  very strong \lyb \ line at $\za = 0.6606$ is at 1703.30~\AA \
whereas \lya \ at the galaxy redshift is expected at 1699.02~\AA \ (see
text), that is 750\kms.
\\ \noindent
$^{\rm e}$ Same as $^{\rm c}$ but for the \lya \ line at $\za = 0.5454$
(see text). \\ \noindent
$^{\rm f}$ Same as $^{\rm c}$ but for the Si\,{\sc ii}($\lambda 1190.42$) line
at
$\za = 0.6606$ (see text).
\end{table*}
\begin{figure*}
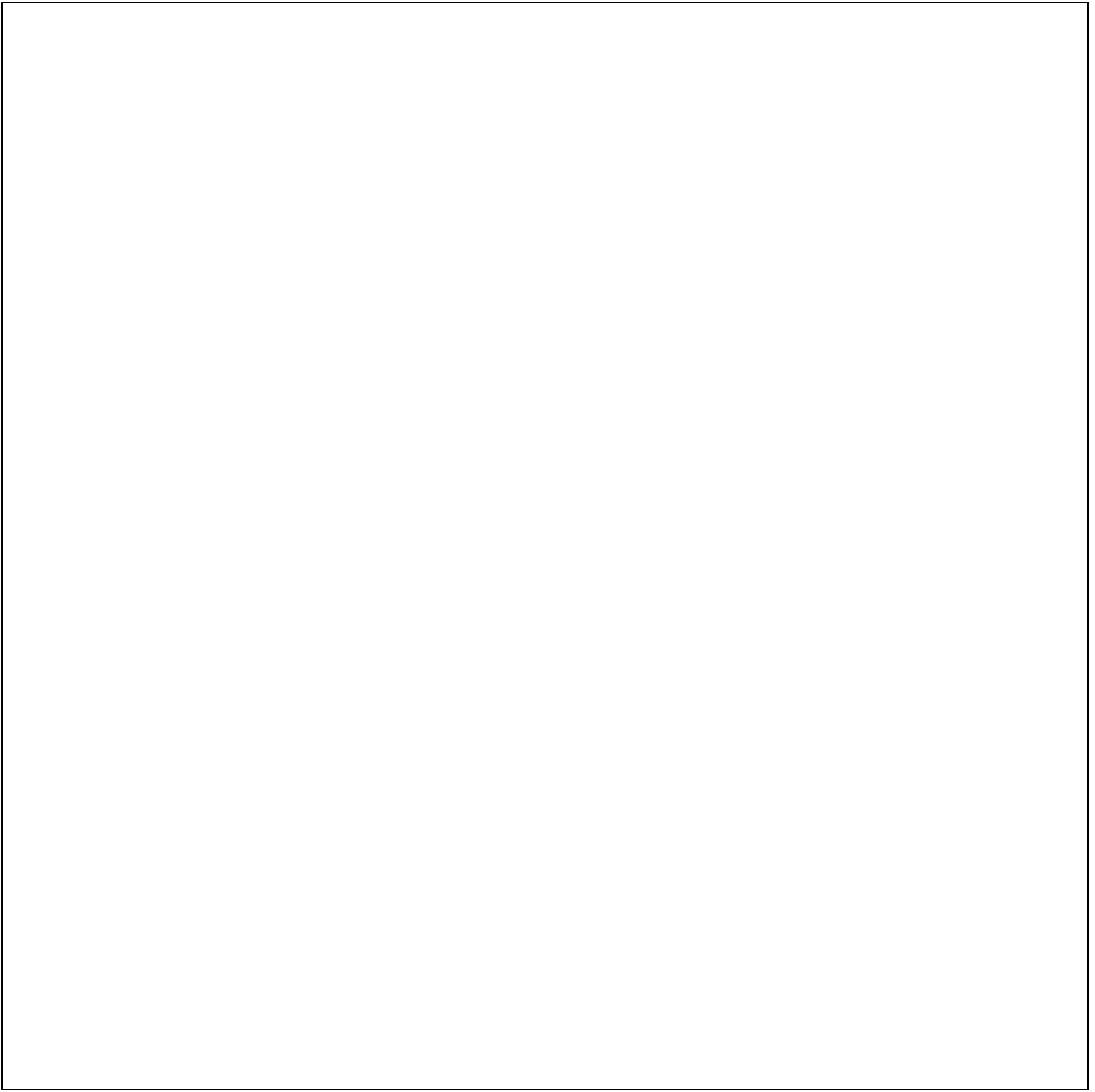

\picplace{18cm}
\caption[]{Field around TON~153. The objects studied spectroscopically are
encircled by the slitlet contours}
\label{Tonpl}
\end{figure*}
\begin{table*}
\caption{Characteristics of the 3C~351 field}
\label{3Ctab}
\begin{tabular}{lrrrrlllll}
\hline\noalign{\smallskip} \\
Gal &$\Delta$x(\arcsec )&$\Delta$y(\arcsec )&$\theta $(\arcsec )&d(kpc)&$\zg$&
$\za$ & Observations &
 m$_{\rm r}$   & M$_{\rm r}$ \\
\noalign{\smallskip}\hline \\
QSO&    0.0  &    0.0  &   0.0 &    0.0 & 0.371   &         &
  & 15.28 &       \\
 1 & +264.5  &  -96.3  & 281.4 &  692.2 & 0.0970  &         & Coinc.$^{\rm a}$
  & 19.8  & -19.1 \\
 2 & +249.7  &  +29.8  & 251.4 & 1549.6 & 0.3174  & 0.3172  & \lya \
  & 20.7  & -21.0 \\
 3 & +229.7  &  -90.2  & 246.7 &        & indef.  &         &
  & 21.1  &       \\
 4 & +195.6  &  -60.2  & 204.7 &  498.9 & 0.0962  &         & Coinc.$^{\rm a}$
  & 16.5  & -22.4 \\
 5 & +155.1  &  -23.8  & 156.9 &  389.3 & 0.0976  &         & Coinc.$^{\rm a}$
  & 18.5  & -20.5 \\
6a & +139.3  &  -85.0  & 163.2 &  934.4 & 0.2840  & 0.2845  & \lya
  & 21.0  & -20.4 \\
6b & +134.0  &  -85.9  & 159.2 &  296.2 & 0.0708  &         & Coinc.$^{\rm b}$
  & 18.2  & -20.0 \\
 7 & +105.0  &  +42.9  & 113.4 &  908.8 & 0.4920  &         & back. gal.
  & 21.5  & -21.3 \\
 8 &  +94.1  &  -15.7  &  95.4 &  860.0 & 0.6217  &         & back. gal.
  & 21.9  & -21.6 \\
10 &  +70.0  &  +13.3  &  71.3 &  347.0 & 0.2260  & 0.2229  & \lya
  & 21.3  & -19.8 \\
11 &  +51.2  &  +17.6  &  54.1 &  125.3 & 0.0910  & 0.0920  & \lya \
  & 18.6  & -20.2 \\
12 &  +31.5  &   -9.7  &  32.9 &  330.6 & 0.7974  &         & back. gal.
  & 21.5  & -22.7 \\
13 &  +14.2  &  -30.4  &  33.6 &   62.9 & 0.0713  &         & Coinc.$^{\rm b}$
  & 21.8  & -18.0 \\
14 &   -2.9  &   +7.0  &   7.6 &        & indef.  &         &
  & 20.7  &       \\
15 &  -26.2  &  +15.3  &  30.2 &        & indef.  &         &
  & 21.1  &       \\
16 &  -39.5  &  +46.2  &  60.6 &        & indef.  &         &
  & 20.3  &       \\
17 &  -54.4  &  +10.2  &  55.1 &  394.4 & 0.4033  &         & back. gal.
  & 21.5  & -20.8 \\
18 &  -73.3  &  + 6.8  &  73.1 &  310.1 & 0.1873  & 0.1880  & \lya \
  & 18.2  & -22.3 \\
19 &  -99.4  & +127.2  & 161.4 & 1440.5 & 0.6070  &         & back. gal.
  & 20.3  & -23.1 \\
20 & -115.0  &  -86.3  & 143.8 &  559.7 & 0.1682  & 0.1676  & \lya \
  & 20.8  & -19.4 \\
21 & -136.2  &   +4.3  & 136.2 &        & indef.  &         &
  & 21.5  &       \\
22 & -148.0  &   -2.3  & 148.0 &  710.6 & 0.2217  & 0.2216, 0.2229  & metal,
\lya
  & 21.4  & -19.5 \\
23 & -166.4  & -152.7  & 225.8 & 1956.2 & 0.5715  &         & back. gal.
  & 20.7  & -22.5 \\
24 & -184.4  & -134.3  & 228.1 &        & indef.  &         &
  & 19.8  &       \\
25 & -209.4  & -118.1  & 240.4 & 1973.8 & 0.5154  &         & back. gal.
  & 19.8  & -23.1 \\
26 & -227.5  & -101.3  & 249.0 &        & indef. &         &
  & 21.4  &        \\
27 & -242.9  &  -41.2  & 246.4 &        & indef. &         &
  & 19.8  &        \\
28 & -261.7  &   11.0  & 261.9 & 1962.7 & 0.4360 &         & back. gal.
  & 20.4  & -22.1  \\
\noalign{\medskip}\hline
\end{tabular}
\\ \smallskip\noindent
$^{\rm a}$ The C\,{\sc ii} galactic line occurs at less than 750\kms \ from the
galaxy redshift (see text). \\ \noindent
$^{\rm b}$  The geocoronal O\,{\sc i} line is at 1302.68~\AA \ whereas the
associated
\lya \ line is expected at 1301.98~\AA \ (see text).
\end{table*}
\begin{figure*}
\picplace{18cm}
\caption[]{Same as Fig.~\ref{Tonpl} for 3C~351}
\label{3Cpl}
\end{figure*}
\begin{table*}
\caption{Characteristics of the H~1821+6419 field}
\label{H18tab}
\begin{tabular}{lrrrrlllll}
\noalign{\smallskip}\hline \\
Gal &$\Delta$x(\arcsec )&$\Delta$y(\arcsec )&$\theta$(\arcsec )&d(kpc)&$\zg$
& $\za$ & Observations & $\mr$   & $\Mr$ \\
\noalign{\medskip}\hline \\
QSO&    0.0 &   0.0 & 0.0   & 0.0    & 0.297  &        &             & 14.24&
       \\
 1 & +273.4 & +85.9 & 286.6 & 692.2  & 0.0000 &        & star        & 20.7 &
       \\
 2 & +255.2 & +47.5 & 259.6 & 1534.3 & 0.2977 &        & QSO cluster & 18.6 &
 -23.0 \\
 3 & +233.3 & +72.0 & 244.2 & 1652.3 & 0.3672 &        & back. gal.  & 19.9 &
 -22.2 \\
 4 & +215.4 & +71.7 & 227.1 & 1925.5 & 0.5482 &        & back. gal.  & 21.7 &
 -21.4 \\
 5 & +192.0 & +83.6 & 209.4 & 1243.5 & 0.2998 &        & QSO cluster & 20.1 &
 -21.5 \\
 6 & +174.4 &  +8.8 & 174.6 & 1016.1 & 0.2909 &        & QSO cluster & 18.9 &
 -22.6 \\
 7 & +148.7 & +24.9 & 150.8 &  889.0 & 0.2966 &        & QSO cluster & 20.1 &
 -21.5 \\
 8 & +132.3 & +50.9 & 141.8 &  556.8 & 0.1703 & 0.1704 & \lya \  & 16.9 &
 -23.3 \\
 9 & +114.1 & +90.3 & 145.5 &  843.5 & 0.2890 &        & QSO cluster & 18.8 &
 -22.7 \\
11 &  +81.9 & +42.5 &  92.2 &  544.0 & 0.2970 &        & QSO cluster & 18.4 &
 -23.2 \\
12 &  +61.6 & -27.1 &  67.3 &  392.6 & 0.2920 &        & QSO cluster & 19.7 &
 -21.8 \\
13 &  +38.9 & +16.1 &  42.1 &  246.6 & 0.2938 &        & QSO cluster & 19.5 &
 -22.0 \\
14 &  +14.2 & -31.3 &  34.4 &  201.1 & 0.293  &        & QSO cluster & 18.8 &
 -22.7 \\
15 &   -4.9 &  +8.1 &   9.5 &   56.6 & 0.3013 &        & QSO cluster & 18.8 &
 -22.8 \\
16 &  -25.1 &  -7.0 &  26.1 &  153.3 & 0.295  & 0.2972 & metal       & 20.5 &
 -21.1 \\
17 &  -47.9 & +16.5 &  50.6 &  289.9 & 0.2844 &        & QSO cluster & 20.7 &
 -20.7 \\
18 &  -61.0 & -45.8 &  76.3 &  455.1 & 0.3018 &        & QSO cluster & 19.8 &
 -21.8 \\
19 &  -87.1 &  -7.0 &  87.4 &  509.5 & 0.2917 &        & QSO cluster & 18.9 &
 -22.6 \\
20 & -106.6 & -21.5 & 108.8 &  641.8 & 0.2969 &        & QSO cluster & 20.7 &
 -20.9 \\
21 & -131.4 &  +3.8 & 131.5 &  781.2 & 0.3000 &        & QSO cluster & 19.5 &
 -22.1 \\
22 & -149.1 & +70.7 & 165.1 &  672.6 & 0.1785 & 0.1806 & \lya \      & 20.3 &
 -21.0 \\
23 & -184.5 &+125.9 & 223.3 &  952.4 & 0.1894 &        &$\Wr \le
0.15$~\AA$^{\rm a}$& 19.3 &
 -22.2 \\
24 & -202.0 &  +3.5 & 202.0 & 1187.3 & 0.2953 &        & QSO cluster & 19.8 &
 -21.7 \\
25 & -225.6 & -61.3 & 233.8 & 1392.0 & 0.301  &        & QSO cluster & 19.4 &
 -22.2 \\
26 & -244.2 & +15.6 & 244.7 & 1456.9 & 0.3010 &        & QSO cluster & 20.0 &
 -21.6 \\
\noalign{\medskip}\hline
\end{tabular}
\smallskip\noindent \\
$^{\rm a}$ This is the $4.5\sigma$ equivalent width limit for an associated
\lya
\ absorption.
\end{table*}
\begin{figure*}
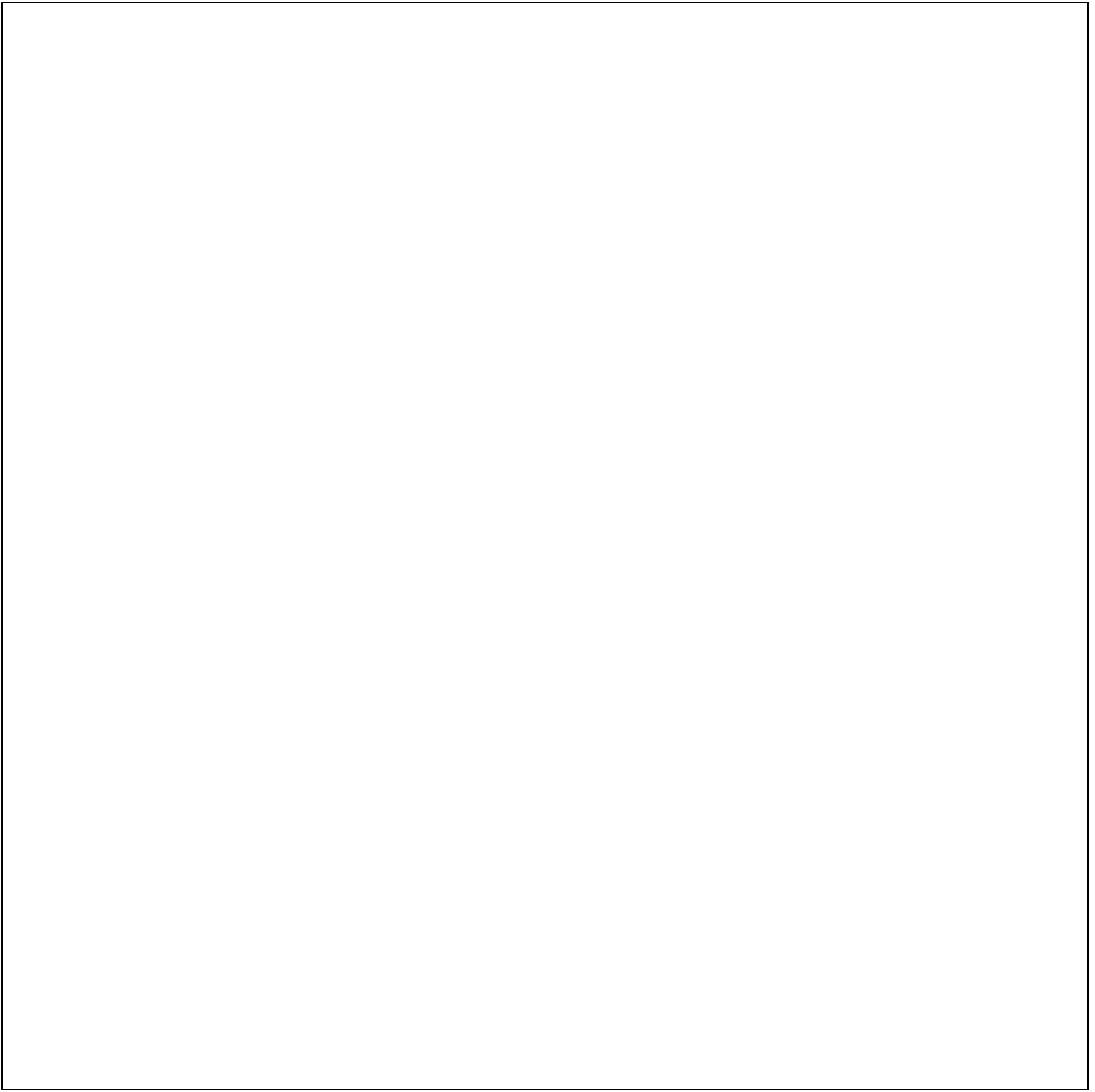

\picplace{18cm}
\caption[]{Same as Fig.~\ref{Tonpl} for H~1821+6419}
\label{H18pl}
\end{figure*}
\subsection{The field toward TON~153 ($z_{\rm em}$=1.022)}
There are 36 \lya -only absorption systems detected in the HST spectrum of
TON~153 (Bahcall et al. 1994). One system is at the quasar redshift and
shows associated O\,{\sc vi} absorption. There are two other metal-rich
systems,
one with only C\,{\sc iv} absorption at $\za=0.2891$ and a very strong
system at $\za=0.6606$ with \lya , \lyb , C\,{\sc ii}, C\,{\sc iii},
Si\,{\sc ii} and Si\,{\sc iii} absorption lines. For the 28 objects in this
field with spectroscopic data, we have obtained 23 redshift identifications,
one of the objects being a star (Table~\ref{Tontab}). The field around TON~153
is shown in Fig.~\ref{Tonpl}, in which the studied objects are encircled by the
slitlet contours. Among the 22 identified galaxies, seven have one or more
associated \lya -only system from the list published by Bahcall et al. (1994).
There is a clump of five \lya \ absorption lines covering an interval of
1400\kms
\ and centred at $\za=0.673$. The metal-rich system, at $\za = 0.6606$, is
within
1600\kms \ of the nearest member of the \lya -only clump (Bahcall et al. 1994).
Three galaxies at $\zg \simeq 0.67$ are clustered in velocity space over
1400\kms \ and, together with the galaxy in the slitlet \#25 ($\zg = 0.6639$),
fall within a projected area of $1.5\times 0.3 h^{-2}_{50}$ Mpc$^{2}$.
Object \# 25 is most probably not responsible for the strong metal-rich system,
given its large impact parameter ($l = 1420\h50$~kpc). Since we do not have a
complete, magnitude limited sample, it is likely that the $z = 0.6606$ absorber
is to be found among the galaxies close to the quasar sightline which have not
been investigated spectroscopically. The same comment applies to the galaxy
in slitlet \#14, which is exactly at the redshift of the $\za = 0.2891$
C\,{\sc iv} absorption system, but at a large impact parameter, $440\h50$~kpc.
\par
Two \lya \ systems are at less than 750\kms \ of each of the two galaxies \#16
and \#19. For the closest to the quasar sightline, we have assigned the closest
system in redshift space.
\par
Two galaxies (\# 6 and \# 12) can a priori be assigned to the same \lya \
absorption line at $\za = 0.6952$, but this line is  clearly a blend of
two components, at $\za = 0.6947$ and 0.6958. Each of them has thus been
assigned to the galaxy with the closest redshift.
\par
For three other galaxies(\#20a, \#20b and \#23), there is no associated \lya \
absorption in the published lists. We thus checked for possible \lya \
absorption at a detection limit less than $4.5 \sigma$. The pair of
interacting galaxies in the slitlet \# 20, $570 \h50$~kpc away on the sky from
the quasar image, and at $\zg = 0.5397$, is at about 1000\kms \ from the
$\za = 0.5454$ \lya \ system. A closer look at the $\za = 0.5454$ \lya \
absorption line clearly reveals two components, one at $\lambda = 1878.32$~\AA
\
and the second at $\lambda = 1876.48$~\AA. This second line corresponds to a
\lya
\ absorption at $\za =0.5436$, or 750\kms \ from the galaxy pair.
\par
\lya \ absorption at the redshift of galaxy \# 23 ($\zg = 0.6265$) is expected
at $\lambda = 1977.29$~\AA, whereas the Si\,{\sc ii}($\lambda 1190$) absorption
from the $\za = 0.6606$ is detected at $\lambda = 1976.74$~\AA). The latter
has an asymmetrical shape, and if the blended component at $\lambda =
1977.7$~\AA
\ is identified as a \lya \ absorption, this leads to a redshift of
$\za = 0.6267$, and thus a relative velocity of 50\kms \ from the galaxy.
We thus add these three absorption lines to the HST \lya \ list, and then
find ten galaxies at less than 750\kms \ from a \lya \ absorption line.
\par
A \lya \  absorption line at $\zg = 0.5674$ (slitlet \#3) would fall at
$\lambda = 1905.4$~\AA, where no line is detected at a $3\sigma$ level.
However,
this is only 650\kms \ away from a line identified as \lyb \ at $\za$ = 0.8544
and quoted as blended in Bahcall et al. (1994). Since we cannot obtain an
unambiguous two component fit, we cannot conclude for this case on the absence
of \lya \ absorption at this galaxy redshift. Similarly, for the galaxy at $\zg
=
0.3976$ (slitlet \#10), there is no absorption at the $3\sigma$ level at the
expected \lya \ wavelength, 1699~\AA. However, the very strong \lyb \ line of
the
$\za = 0.6606$ metal-rich system lies 500\kms \ redward to the latter
wavelength,
and we also cannot conclude for this case.
\par
For the 8 remaining galaxies with identified redshifts, the wavelength
range of the possible, associated \lya \ absorption has not yet been observed
with the HST.
\par
The completeness of the survey is given in Columns 2 to 4 of
Table~\ref{galpop}.
All the galaxies with associated  \lya -only and metal-rich absorption systems
are intrinsically bright ($M_{\rm r} < -21.1$).
\subsection{The field toward 3C~351 ($z_{\rm em}$=0.371)}
There are 14 \lya -only absorption lines listed in Bahcall et al. (1993), of
which two are within 3000\kms \ of the quasar emission redshift. Furthermore,
there are two metal-rich systems at $\za = 0.2216$ and $\za = 0.3646$. We have
obtained spectra for 28 objects in this field, shown in Fig.~\ref{3Cpl}, and
get
20 successful identifications, all being galaxies. This field has
also been studied by LBTW who identified 11
galaxies, of which six (our slitlets \# 5, \# 6, \# 11, \# 12, \# 16 and
\# 18, respectively \# 261, \# 245, \# 187, \# 180, \# 121 and \#  92 in their
numbering), are common to our sample.
There is a good agreement in the $z$ determination ($\Delta z \le 0.001$) for
galaxies \#5, \# 6 and \# 18, and a small difference for \# 11 at $\zg =
0.0910$,
for which they measure $\zg = 0.0920$ . We did not succeed in measuring
a redshift for galaxy \# 16, for which LBTW obtain $\zg =
0.2566$.
Furthermore, we disagree on
the redshift of the galaxy \# 12 (\# 180 in LBTW). Their spectrum
shows two emission lines, around 5000 and 6700~\AA, that they identify as
 [O\,{\sc ii}]$\lambda 3727$ and [O\,{\sc iii}]$\lambda 5007$ respectively.
  They do not detect [O\,{\sc iii}]$\lambda 4958$,
which one would expect given the strong [O\,{\sc iii}]$\lambda 5007$ line (see
their Fig.~14). We do not detect their [O\,{\sc ii}] line around 5000~\AA, but
we
identify the line at 6698.89~\AA \ with [O\,{\sc ii}], since we also detect the
Ca\,{\sc ii} absorption doublet. We then get a redshift of 0.798 for this
galaxy.
\par
Among our 20 identified galaxies, seven are at less than 750\kms \ of a \lya \
absorption line of the HST complete sample, including \# 22 which is at a
redshift close that of the $\za = 0.2216$ metal-rich system, but with a very
large impact
parameter ($l = 710\h50$~kpc). Three other have nearly the same redshift $\zg
\simeq 0.097$, and one then expects an associated \lya \ line at $\lambda =
1333.59$~\AA. No line is present at this wavelength in the HST sample, but the
C\,{\sc ii}($\lambda 1134$) Galactic line is detected only 600\kms \ away.
There is an indication of an asymmetrical shape for this line, but
only high resolution spectroscopy could confirm this. The same problem occurs
for
the two galaxies \# 6b and \# 13 at $\zg \simeq 0.071$, for which an associated
\lya \ line would be at $\lambda = 1301.98$~\AA, close to the very strong
geocoronal O\,{\sc i} emission line at $\lambda = 1302.68$~\AA. Thus, no
conclusion can be reached in this case either. The eight remaining objects are
background galaxies. We do not observe any unambiguous case of a galaxy which
does not give rise to a \lya \ absorption line. Information on the galaxy
counts
in different magnitude bins is given in Columns 5 to 7 of Table~\ref{galpop}.
The
completeness is calculated including the redshifts obtained by LBTW, which
correspond to the numbers in bracket in Column~6. When a galaxy is common to
the
two samples, we adopt our own redshift determination.
\par
In this field, the galaxies with a redshift close to that
of a \lya \ system  are less luminous  in average than in
the field of TON~153 ($M_{\rm r} < -19.4$).
\subsection{The field toward H~1821+6419 ($z_{\rm em}$=0.297)}
The HST complete sample of absorption lines contains 12 \lya -only absorption
systems and one high-ionization metal-rich system with C\,{\sc iv} and Si\,{\sc
iv} absorption lines at the quasar redshift. In this field (see
Fig.~\ref{H18pl})
we measured the redshifts of the 25 objects which were in the slitlets,
all being fairly bright ($m_{\rm r} \le 21.0$). The sample contains
one star and 24 galaxies, of which two are background galaxies, two are at less
than 650\kms \ of a \lya \ absorption line, and 19 other galaxies are within
4000\kms \ of the quasar redshift. They confirm the presence of a cluster
at $\za \simeq z_{\rm e}=0.2972$, already inferred by Schneider et al. (1992).
{}From
these 19 galaxies, we get a value of 1100\kms \ for the velocity dispersion.
\par
The remaining galaxy has a redshift of $\zg = 0.1894$, and there is no \lya \
absorption line detected in the quasar spectrum at this redshift. The
$4.5\sigma$ rest equivalent width limit is 0.15~\AA, and the nearest \lya \
system is 2200\kms \ away.
\par
Schneider et al. (1992) have also studied this field and measured the redshifts
of
eight galaxies, none of them being common to our sample. They found 6 galaxies
at $\zg \simeq z_{\rm e}$ , one background galaxy and one at a redshift close
to
that of a \lya \ system at $\za=0.2265$. The completeness of the redshift
survey
(Columns 8 to 10 of Table~\ref{galpop}) is calculated including their redshift
measurements, which are given in brackets in Column~9.
\par
The nebulosity North-West to the quasar is the planetary nebula Kohoutec~1-16.
\subsection{Overall results}
We have obtained the spectra of 81 objects for the three combined fields. We
identified two stars and could determine 66 galaxy redshifts, thus
a success rate of 81\%. The very small number of stars in our sample is a
consequence of having selected resolved objects. Since the selection was made
using high resolution imaging data (0.6\arcsec \ FWHM) of the central parts of
the fields, we were able to include compact galaxies in our sample. Among the
66
galaxies with measured redshift, 19 (16) are at less than 750
(500)\kms \ of a \lya \ absorption system.
\par
Of the remaining galaxies, only one clearly does not have an associated \lya \
absorption, whereas seven galaxies have an expected, associated \lya \
absorption possibly blended with either a geocoronal line, a Galactic line or a
metal-line from another system. Higher resolution data are needed to draw
any conclusion on these seven cases. The relative velocity between four
galaxies
and a metal-rich system is less than 600\kms ; one of these galaxies
belong to the above sample of \lya -only-galaxy associations. There are 18
galaxies from the cluster associated with the quasar H~1821+6419 and ten
background
galaxies. Finally, there are eight galaxies, all in the field of TON~153, for
which there is no available data in the wavelength range of the expected,
associated \lya \ absorption.
\section{Properties of \lya \ absorbers}
\subsection{The samples}
%
\subsubsection{The \lya \ sample}
The rest-frame equivalent width limit for the HST spectra of 3C~273 are much
lower than those obtained for the other quasars discussed here or by LBTW.
We thus select \lya \ systems with $\Wr \left({\rm Ly}\alpha \right)
\ge 0.10$~\AA, limit
consistent with the weakest systems detected in the quasar spectra of our
sample.
\par
The \lya \ absorption lines detected only in low-resolution FOS spectra are
excluded from the combined sample. This resolution, $R=180$, is seven times
lower
than that obtained with the higher resolution mode, and very strong lines could
be blends of several lines or damped  \lya \ lines from a metal-rich system.
Indeed, lines as strong as $\Wr \ge 1.5$~\AA \ detected in the $R=1300$
resolution mode are all associated with metal-rich systems and are not present
in
the homogeneous sample of 109 \lya -only systems build from the 17 Key Project
quasar spectra (Bahcall et al. 1994). Since metal-rich systems are known to be
physically associated with galaxies, they are not included in the \lya \
sample,
as this could introduce a bias in the statistical analyses of the \lya \
absorber
population.
\subsubsection{The galaxy and the \lya \ absorber-galaxy association samples}
Three other teams obtained galaxy redshift samples for HST quasar fields.
LBTW have studied 8 fields and their relative velocity
criterion for assigning absorption systems to galaxies is $|\Delta v| \le 1000
$\kms. Their sample remains unmodified if our criterion ($|\Delta v| \le
750$\kms) is applied instead, since it does not contain any association with an
absolute
relative velocity in the range 750-1000\kms.  Morris et al. (1993) have
obtained
complete, magnitude limited samples of over 200 galaxies in the field of
3C~273. These samples were defined to carry out a correlation analysis and are
complete to faint magnitude limit out to a large impact parameter of
$16\h50$~Mpc. For their sample of 17 \lya \ absorber-galaxy associations, the
largest velocity
separation is 710\kms, but only 120\kms \ for impact parameters smaller than
$2.5\h50$~Mpc. The small sample studied by Schneider et al. (1992) contains
only
one galaxy at $\zg \simeq \za \left({\rm Ly}\alpha \right) \neq z_{\rm e}$ and
the relative velocity is 340\kms. Consequently our relative velocity selection
criterion does not exclude any galaxy from the samples obtained by the above
mentioned teams. For the cases in common with LBTW, we use our
own redshift estimates.
\par
The criteria for assigning a galaxy to a \lya \ absorption line are the
smallest
 relative velocity and impact
parameter. Even though the galaxy sample is far from being complete, the
additional second criterion is needed in two cases: the close pair at $\zg =
0.5397 \simeq \za = 0.5436$ in the field of TON~153 and the two galaxies at
$\zg
\simeq \za = 0.2229$ in the field of 3C~351. This leads to a total sample of 32
\lya \ absorber-galaxy associations, of which two are in common with the sample
of LBTW, and to a sample of 11 galaxies without associated \lya \ absorption.
The
combined sample is given in Table~6.
\begin{table*}
\label{all_abs}
\caption{The combined sample of identified \lya \ systems}
\begin{tabular}{lllrlrlr}
\noalign{\smallskip}\hline \\
Quasar & Author & z$_{\rm a}$ & w$_{\rm rest}$ & z$_{\rm g}$ &
d(kpc) & M$_{\rm r}$ & $\Delta v$ \\
\noalign{\smallskip}\hline \\
 TON~153 (1317+2743) & This work.&&&&&& \\
&&0.5349 &  0.29 &  0.5324 & 1624.0 & -21.3 &  490  \\
&&0.5436 &  0.10 &  0.5397 &  563.5 & -22.3 &  750  \\
&&0.6267 &  0.11 &  0.6265 & 1421.1 & -22.8 &   40  \\
&&0.6692 &  0.60 &  0.668  & 1242.4 & -23.2 &  220  \\
&&0.6716 &  0.39 &  0.6715 &   77.5 & -22.0 &   20  \\
&&0.6736 &  0.78 &  0.6717 &  415.3 & -21.3 &  340  \\
&&0.6947 &  0.34 &  0.6952 & 1669.7 & -22.0 &  -90  \\
&&0.6958 &  0.40 &  0.6974 &  996.9 & -21.4 & -280  \\
&&0.7667 &  0.12 &  0.7663 & 2375.7 & -21.3 &   70  \\
 3C~351 (1704+6048) & This work.&&&&&& \\
&&0.0920 &  0.82 &  0.0910 &  125.3 & -20.2 &  280  \\
&&0.1676 &  0.23 &  0.1682 &  559.7 & -19.4 & -150  \\
&&0.1880 &  0.36 &  0.1873 &  310.1 & -22.3 &  180  \\
&&0.2229 &  0.34 &  0.2260 &  347.0 & -19.8 & -750  \\
&&0.2845 &  0.62 &  0.2840 &  934.4 & -20.4 &  120  \\
&&0.3172 &  0.64 &  0.3174 & 1549.6 & -21.0 &  -50  \\
 3C~351 (1704+6048) & Lanzetta et al.$^{\rm a}$&&&&&& \\
&&0.0920 &  0.82 &  0.0922 &  125.3 & -20.2 &  280  \\
&&0.1611 &  0.29 &  0.1621 &  718.8 & -22.9 & -260  \\
&&0.1880 &  0.36 &  0.1875 &  310.1 & -22.3 &  130  \\
&&0.3621 &  0.49 &  0.3621 &  189.2 & -21.5 &    0  \\
&&0.3716 &  0.26 &  0.3709 &  301.6 & -23.7 &  150  \\
&&       &$<0.16$&  0.0860 &  325.3 & -21.0 &      \\
H~1821+6419 & This work  &&&&&& \\
&&0.1704 &  0.64 &  0.1703 &  556.8 & -23.3 &   30  \\
&&0.1806 &  0.30 &  0.1785 &  672.6 & -21.0 &  530  \\
&&       &$<0.15$&  0.1894 &  952.4 & -22.2 &      \\
H~1821+6419 & Schneider et al. &&&&&& \\
&&0.2251 &  0.75 &  0.2265 &  158.2 & -22.1 & -340  \\
1354+1933 & Lanzetta et al.$^{\rm a}$ &&&&&& \\
&&0.4306 &  1.03 &  0.4302 &  210.5 & -20.5 &   80  \\
&&       &$<0.25$&  0.3509 &  543.9 & -20.6 &      \\
&&       &$<0.15$&  0.4406 &  187.0 & -21.3 &      \\
&&       &$<0.15$&  0.5293 &  415.6 & -21.4 &      \\
TON~28 (1001+2910) & Lanzetta et al.$^{\rm a}$ &&&&&& \\
&&0.2132 &  0.57 &  0.2127 &  341.8 & -22.4 &  120  \\
&&       &$<0.08$&  0.1343 &  247.3 & -21.2 &      \\
US~1867 (0850+4400) & Lanzetta et al.$^{\rm a}$ &&&&&& \\
&&0.4435 &  0.51 &  0.4428 &   77.9 & -21.8 &  150  \\
&&       &$<0.24$&  0.4259 &  380.8 & -21.4 &      \\
&&       &$<0.08$&  0.5065 &  210.0 & -21.9 &      \\
&&       &$<0.09$&  0.5200 &  375.5 & -23.0 &      \\
3C~95 (0349$-$1438)  & Lanzetta et al.$^{\rm a}$ &&&&&& \\
&&       &$<0.17$&  0.4223 &  584.9 & -22.0 &      \\
&&       &$<0.11$&  0.5526 &  602.8 & -22.2 &      \\
3C~273 (1226+0219) & Morris et al.$^{\rm b}$ &&&&&& \\
&&0.0034 &  0.37 & 0.00333 & 1376.0 & -18.6 &   20  \\
&&0.0053 &  0.41 & 0.00524 &  368.0 & -17.3 &   20  \\
&&0.0073 &  0.24 & 0.00745 &  944.0 & -16.2 &  -50  \\
&&0.0294 &  0.12 & 0.02938 &  560.0 & -18.1 &   20  \\
&&0.0490 &  0.14 & 0.04957 & 2592.0 & -18.8 & -160  \\
&&0.0666 &  0.30 & 0.06892 & 7600.0 & -20.0 & -650  \\
&&0.1201 &  0.11 & 0.12067 & 3056.0 & -20.2 & -180  \\
&&0.1466 &  0.29 & 0.14668 & 1600.0 & -21.3 &  -20  \\
\noalign{\medskip}\hline
\end{tabular}
\smallskip\noindent\\
$^{\rm a}$ The sample of LBTW has been reduced to the high
resolution data, excluding the G160L data.
\noindent\\
$^{\rm b}$ The sample of Morris et al. (1993) is restricted to
$\Wr \left({\rm Ly}\alpha \right) \ge 0.10$~\AA.
\end{table*}
\subsection{Significance of the \lya \ absorber-galaxy associations}
In discussing the relation between the galaxies found in the field of each
quasar and \lya \ lines, it is important to assess the significance level
of the ``associations'' defined above. In other words, can we rule out that
they
are just due to chance? First of all, one has to decide in which manner the
distance between the absorber and the ``associated'' galaxy is to be
estimated. As discussed by Morris et al. (1993), one should compute
the 3-dimensional distance which involves both the separation in depth (related
to the redshift difference) and the impact parameter. However, it is
straightforward to verify (for instance using the formulas given by Tytler et
al. (1993) for $q_0=0.5$) that the contribution of the ``transverse'' term can
be neglected. Indeed, for a $\Delta z$ value corresponding to
$\Delta v=100$\kms \ and
an angular separation $\Delta \theta = 3\arcmin$ at $z=0.5$, the latter barely
equals the separation in depth. Then, for simplicity we consider the comoving
distance in depth only, which for small redshift differences reduces to
\begin{equation}
 d = \left( {c \over H_0}\right) {| \Delta z| \over \left( 1+z\right) }.
\end{equation}
Since our galaxy sample is not complete, we first consider  the detected
galaxies, and for each of them search for the nearest \lya \ line.
In order to get homogeneous results for our three quasar fields, we
restrict our search to \lya \ lines with a rest-frame equivalent width
$w_{\rm r} \geq 0.24$~\AA, and thus consider only the systems included in the
complete samples of the HST absorption line survey (Bahcall et al. 1993, 1994)
(the deblended lines in the
spectrum of TON~153 quoted in Table~\ref{Tontab} are not considered).
Given the differences in S/N ratios and redshift ranges, such lines can be
detected over the redshift intervals 0.39-1.00, 0.05-0.357 and 0.0-0.285 in
TON~153, 3C~351 and H~1821+6419 respectively. The same redshift intervals are
used for the galaxy sample. As a measure of the overall match between the sets
of redshifts of \lya \ absorbers and galaxies,
Morris et al. (1993) consider the average of the nearest distance
over all galaxies. However, this quantity gives much weight to a single
galaxy far from any \lya \ line even if several close associations are present,
i.e. few anti-coincidences may hide several significant coincidences. We thus
prefer instead to count the number of ``associations'' found  for the following
relative velocity criteria: $|\Delta v| < 750$\kms , the value adopted
above, $|\Delta v| < 400$\kms , which is roughly the cutoff in the relative
velocity distribution (see Sect.~4.1) and $|\Delta v |
< 250$\kms, which is about the average galactic inner velocity dispersion.
\begin{table*}
\caption{Significance of the \lya \ absorber-galaxy associations}
\label{assoc}
\begin{tabular}{llllllllll}
\hline\noalign \\
Quasar field     & $z$ interval& N$_{\rm gal}$ & N$_{\rm syst}$ &
N$_{1,{\rm obs}}^{\rm a}$ & N$_{2,{\rm obs}}^{\rm a}$& N$_{3,{\rm obs}}^{\rm
a}$&
P$_{1}^{\rm b}$           & P$_{2}^{\rm b}$          & P$_{3}^{\rm b}$ \\
\noalign{\hrule} \\
TON~153    & 0.39 - 1.000 & 13 &  26  &  4 & 5 & 6 & 0.059 & 0.062  & 0.18   \\
3C~351     & 0.05 - 0.357 &  7 &  12  &  4 & 6 & 6 & 0.001 & 5~10$^{-4}$ &
5~10$^{-4}$ \\
H~1821+6419 & 0.0  - 0.285 &  3 &  11  &  1 & 1 & 2 & 0.22  & 0.33   & 0.11
\\
combined     &              & 23 &  49  &  9 &12 &14 &5~10$^{-5}$ & 5~10$^{-5}$
&
                                                                    5~10$^{-5}$
\\
\noalign{\smallskip}\hline
\end{tabular}
\smallskip\noindent \\
$^{\rm a}$ $N_{1,{\rm obs}}$, $N_{2,{\rm obs}}$, $N_{3,{\rm obs}}$ are the
observed numbers of \lya \ absorber-galaxy pairs with  $|\Delta v|
\le 250$, 400, 750\kms respectively \noindent \\
$^{\rm b}$ $P_{1}$, $P_{2}$, $P_{3}$  are the probabilities to have by chance
at least $N_{1,{\rm obs}}$, $N_{2,{\rm obs}}$ or $N_{3,{\rm obs}}$
\lya \ absorber-galaxy pairs with $|\Delta v| < 250$, 400, 750\kms \
respectively
\end{table*}

To judge whether the observed number of redshift agreements is significant
or not, we perform Monte Carlo simulations. The same redshift intervals are
used
together with the same galaxy redshifts for each quasar field. Next, \lya \
redshifts are selected at random in the appropriate interval and with a mean
distribution of the form  (Bahcall et al. 1994):
\begin{equation}
{{\rm d}N \over {\rm d}z} \propto \left( 1+z\right) ^{0.5},
\end{equation}
with the same number of lines than the observed one. Each simulation is then
analyzed in the same way as the original data (search for the nearest line to
each galaxy) and we thus derive the probability to get by  chance a number of
associations which is equal to the observed one or greater. This is done first
for the three quasar fields separately and then globally. For the latter case,
a simulation consists in joining three individual ones for each of the three
quasar fields and in considering the total number of associations. Results are
given in Table~\ref{assoc}. The associations are not statistically significant
for individual fields, except for 3C~351, but for the combined sample the
existence of redshift agreements is found to be highly significant.
\subsection{Search for correlations}
\begin{figure}
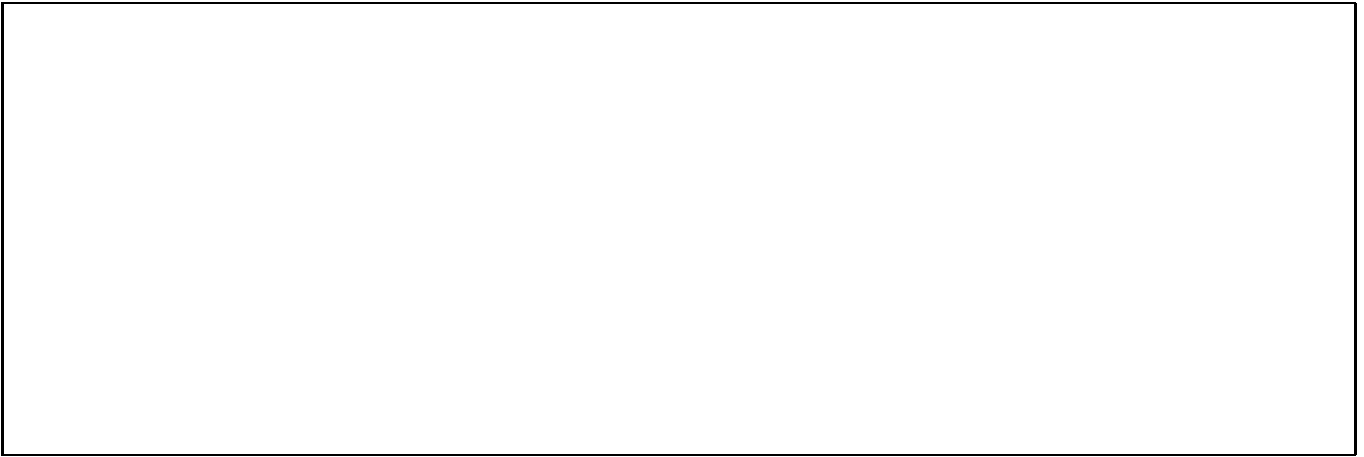

\picplace{6cm}
\caption[]{Impact parameter versus \lya \ rest-frame equivalent width.
Crosses: This work, Hexagons: Lanzetta et al. (1995), Stars:
Morris et al. (1993), Diamonds: Schneider et al. (1992)}
\label{DW}
\end{figure}
%
\subsubsection{Impact parameter and rest-frame equivalent width}
LBTW have found a statistically significant
anti-correlation between the \lya \ rest-frame equivalent width and the
galaxy impact parameter (see their Fig.~24).
On the basis of their results, they conclude that most luminous galaxies are
surrounded by extended gaseous envelopes of radius $R \simeq 345\h50$~kpc \
 and of roughly unit coverage factor. Their sample includes 11
\lya \ absorber-galaxy associations of which two are metal-rich absorbers and
one
is a very strong \lya \ absorber detected only in low spectral resolution data.
\par
If all the \lya \ absorbers were a single population, analyses of subsamples
selected either by the strength of \lya \ absorption or the metal content
should lead to similar correlations. Since the impact parameters of strong
metal-rich absorbers are known to be smaller than those for \lya \ clouds,
it is appropriate to search independently for correlations among the
``\lya-only'' population.
To investigate the  $\Wr \left({\rm Ly}\alpha \right)$ -$D$ relation, we have
first considered our homogeneous sample which excludes metal-rich systems.
We do not confirm the anti-correlation previously found (see Fig.~\ref{DW}).
The results of the generalized Spearman and Kendall rank-correlation tests are
 given in Table~\ref{corr}.
The parameters $\alpha _{\rm s}$ and $\alpha _{\rm k}$ are
the probabilities that the values of the coefficients obtained for our
distribution, $r_{\rm s}$ and $r_{\rm k}$, are reached if the data are
uncorrelated. As LBTW, we have included the upper limits. We  have used  the
method  presented in Isobe et al. (1986) and implemented in the software
package ASURV Rev1.1 (La Valley et al. 1992).
If the sample is restricted to $\Wr \geq$ 0.24~\AA, criterion limit selected
by Bahcall et al. (1993, 1994) for complete \lya \ samples, there is a
weak anti-correlation between $D$ and $\Wr \left({\rm Ly}\alpha \right)$ at the
90\%  confidence level.
Adding to   our whole combined sample the previously excluded  three strong
\lya \
systems of the Lanzetta et al.' survey strongly modifies the results  of the
rank statistics, the anti-correlation being then present at the
97\%  confidence level.
\par
Stocke et al. (1995) have recently found eight new low-redshift \lya-only
systems,
of which seven are absorber-galaxy associations. As in the case of Morris et
al.'
survey, all the galaxies are at large impact parameters ($\geq 675 \h50$~kpc)
and
they conclude that none of the absorbers is close enough to an individual
galaxy
to be unambiguously associated with it. They have combined their results with
previously published data, including only absorber-galaxy associations.
Analysis of
their sample shows an anti-correlation between $D$ and
$\Wr \left({\rm Ly}\alpha \right)$, which is not as tight when only low
equivalent
widths ($\Wr <$ 0.30~\AA) are considered. This trend
 is similar to that found for our sample when dividing it into strong and weak
lines.
\par
It should be noted that excluding all the censored cases with only
 $\Wr \left({\rm Ly}\alpha \right)$ upper limits could lead to results which
strongly differ from those obtained with complete samples. Ordinary rank
statistics
applied to the 32 absorber-galaxy associations of our combined sample reveals a
strong
anti-correlation   between $D$ and  $\Wr \left({\rm Ly}\alpha \right)$ at the
99.5\%  confidence level, which however is not as significant (95\%  confidence
level) at high equivalent widths, $\Wr \geq$ 0.24~\AA.
Adding to our 32 associations, the three strong \lya \ absorbers of Lanzetta et
al.' survey, leads to a very strong  anti-correlation  (99.98\%  confidence
level).
  \par
The differences in the results found with ordinary and generalized rank
statistics call for caution in drawing conclusions. There appears to be a
weak anti-correlation  between $D$ and $\Wr \left({\rm Ly}\alpha \right)$,
which is of higher significance when either metal-rich \lya \ systems  or
stronger \lya \ absorbers are considered for complete samples with the
 $\Wr \left({\rm Ly}\alpha \right)$ upper limits included. These results do
not support the single population hypothesis.
\par
The \lya \ absorber-galaxy associations with smaller impact parameters
could trace extended galactic halos. Indeed,  all the four galaxies of the
combined sample with impact parameters $D \le 175 \h50$~kpc have associated
\lya -only absorption, whereas this is also the case for only about half of
them for $175 < D \le 1000$~$\h50$kpc. These results are not significantly
modified if  the combined sample is restricted to galaxies brighter than an
absolute magnitude limit, $\Mr = -19$, which is the magnitude limit of our
survey at $z \simeq$ 0.3. This selection criterion excludes the galaxies
at $z < 0.06$ of the survey of Morris et al. (1993). It should be noted
that all the six galaxies at $1 < D < 2$~$\h50$Mpc and brighter than
$\Mr = -19$  have an associated \lya \ absorption.
Consequently, we detect a break in the number of \lya \ absorber-galaxy
associations relative to the total number of galaxies at $D \simeq 200
\h50$ kpc, with a flattening at larger values of $D$ (see also Sect. 5).
This critical value of $D$, which should give the average size of
galactic halos, is nearly twice as small as that found by LBTW.
This difference mainly arises from having included in our sample a large
number of absorber-galaxy associations detected either in our survey or that of
Morris et al. (1993).
\begin{table}
\caption{Correlations}
\label{corr}
\begin{tabular}{lllll}
\hline\noalign \\
Variables  & r$_{\rm s}$ & $\alpha _{\rm s}$ & r$_{\rm k}$ & $\alpha _{\rm k}$
\\
\noalign{\hrule} \\
$D$         , $w_{\rm r}$            & -0.141 & 0.359 & 1.034 & 0.301 \\
$D$         , $w_{\rm r}$ $^{\rm a}$ & -0.336 & 0.087 & 1.650 & 0.099 \\
$M_{\rm r}$ , $\log \left(D\right)$  & -0.243 & 0.115 & 1.527 & 0.127 \\
$w_{\rm r}$ , $|\Delta v|$           & -0.034 & 0.851 & 0.098 & 0.922 \\
$D$         , $|\Delta v|$           &  0.124 & 0.489 & 0.716 & 0.474 \\
\noalign{\smallskip}\hline
\end{tabular}
\smallskip\noindent \\
$^{\rm a}$ Sample restricted to $\Wr \left({\rm Ly}\alpha \right) \ge 0.24$~\AA
\end{table}
\subsubsection{Relative velocity distribution}
\begin{figure}
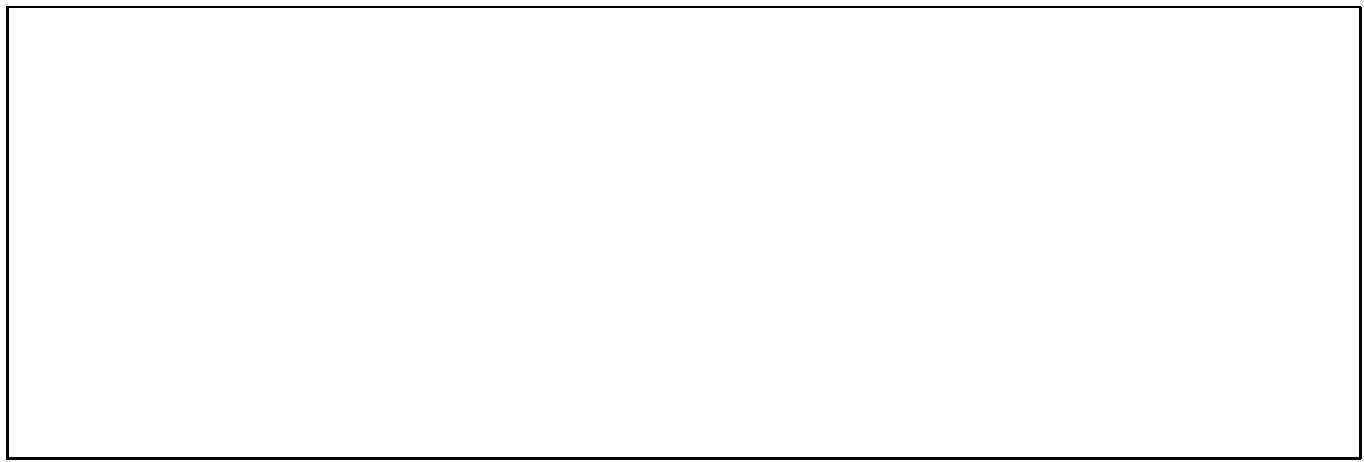

\picplace{6cm}
\caption[]{Relative velocity distribution}
\label{dv_hist}
\end{figure}

Although the adopted relative velocity criterion for assigning a galaxy to a
\lya \ absorption system is $|\Delta v| \le$ 750 \kms, the relative velocity
distribution of \lya \ absorber-galaxy associations, as shown on
Fig.~\ref{dv_hist}, is narrow, roughly centred on $\Delta v = 0$\kms \ and
symmetrical, whereas the expected histogram for a random distribution is flat.
The half-width, half maximum of this distribution is
HWHM $\simeq 120$\kms and 77\% of the 32 \lya \ absorber-galaxy associations
have
$|\Delta v| \le 300$\kms.
\par
This narrow velocity distribution is compatible with spiral galaxy
rotation velocities. Its mean is smaller than  the median velocity  dispersion
of
$\sim 500$\kms \ which characterizes the thickness of the sheets of large-scale
structures (Ramella et al. 1989, 1992). This velocity dispersion has been
derived
by averaging within cells of sizes $\sim 10 \h50$ Mpc, i.e. over one
order of magnitude larger than the typical projected distances between
\lya \ absorber-galaxy associations. The local thickness of large-scale
structures can be derived from the CFA redshift survey of close galaxy-pairs
and it is about a factor of three smaller than the above median velocity
dispersion (M. Geller, private communication). The small HWHM of the relative
velocity distribution of the \lya \ absorber-galaxy associations is
thus also consistent with the assumption that most \lya \ absorbers  trace
large-scale structures rather than individual objects, at least for impact
parameters $D >$ 200 to 300 $\h50$~kpc.
\subsubsection{Impact parameter and luminosity }
\begin{figure}
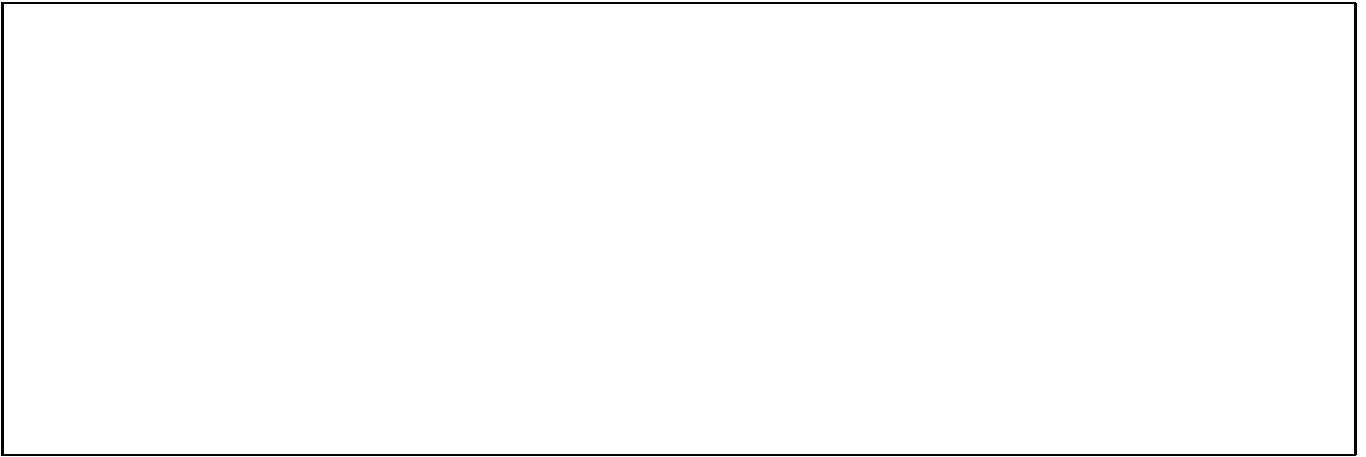

\picplace{6cm}
\caption[]{Luminosity of the galaxy versus impact parameter, and same
notations as in Fig.~\ref{DW}}
\label{MD}
\end{figure}

For the metal-rich absorption systems, the scaling law between the
stellar luminosity and the galactic halo radius follows a power law, $D
\propto L^{\alpha}$, with $\alpha$ = 0.2  to 0.3 (Bergeron \& Boiss\'e
1991; Steidel 1993). This correlation is not as strong as
that found for galactic stellar component for which the Schwarzschild scaling
law implies $\alpha \simeq 0.42$. We investigate the existence of such a
correlation for the \lya \ absorber-galaxy associations. Results are given in
Fig.~\ref{MD} and Table~7, from which it is clear that the impact parameter and
the galaxy luminosity are uncorrelated.
\par
There is an overlap in impact parameters between the
metal-rich and \lya -only absorbers. As the \lya \ absorbers of smallest impact
parameters have larger rest-frame \lya \ equivalent widths, the continuity in
sizes and H\,{\sc i} column densities between the two populations strengthens
the assumption that at least the strongest \lya \ absorbers trace the
outermost parts of galactic halos.  The lack of correlation between the impact
parameter and the luminosity, even for $D < 300 \h50$ kpc, may suggest that the
size of the \lya \ galactic halos is linked to the total mass, which at large
galactic radii is dominated by the dark matter, whereas at smaller radii both
the
baryonic (mainly in the form of stars) and dark matter components contribute to
the total mass.
\begin{figure}
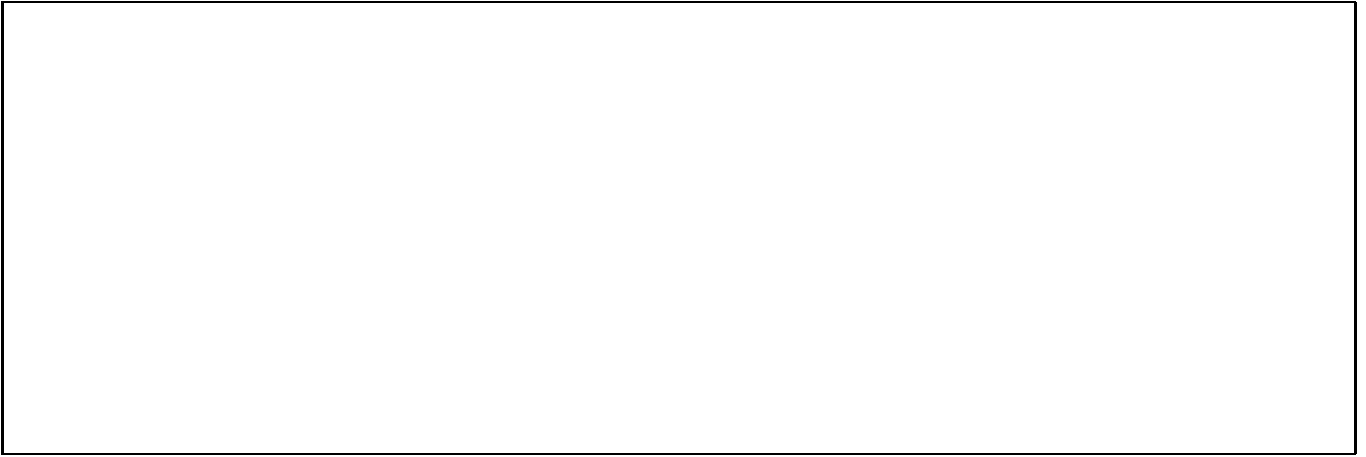

\picplace{6cm}
\caption[]{Impact parameter versus \lya \ absorber-galaxy relative velocity,
and same notations as in Fig.~\ref{DW} }
\label{DV}
\end{figure}
\begin{figure}
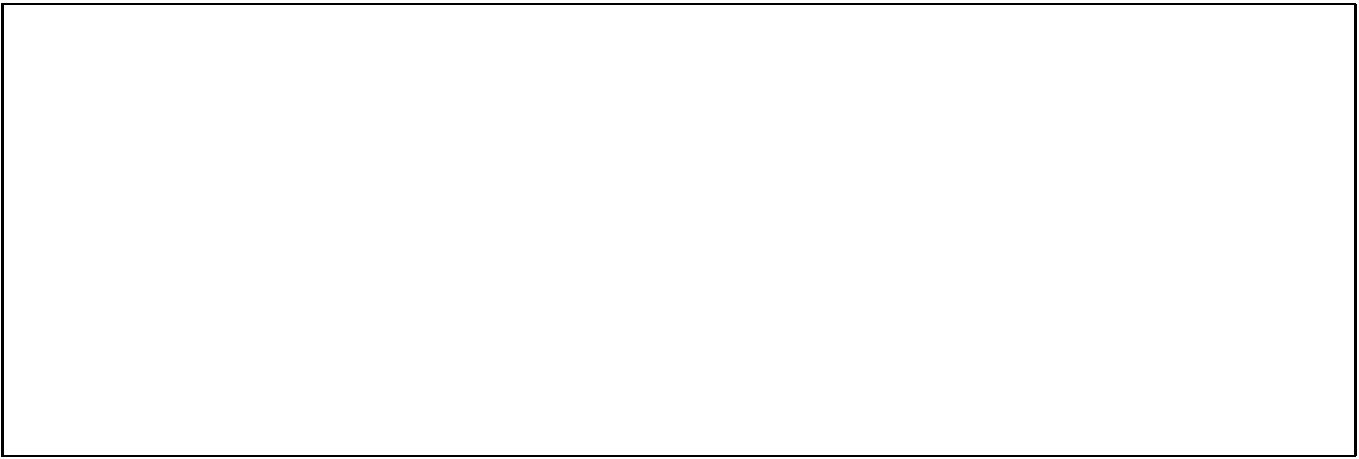

\picplace{6cm}
\caption[]{\lya \ rest-frame equivalent width versus \lya \ absorber-galaxy
relative velocity, and same notations as in Fig.~\ref{DW}}
\label{WV}
\end{figure}
%
\subsubsection{Search for other correlations}
There is no correlation between  the  \lya \ absorber-galaxy  relative
velocity and the impact parameter or the \lya \ rest-frame equivalent width
(see Table~7 and Figs.~\ref{DV} and \ref{WV}). The former suggests a lack of
organized motions on scales 0.3-1.5 $\h50$ Mpc, consistent with the
assumption that a large fraction of the \lya \ absorber-galaxy associations
trace coherent large-scale structures rather than individual galaxies.
\subsection{Clustering of \lya \ systems and galaxies}
The complete galaxy survey of Morris et al. (1993) towards 3C~273 has allowed a
study of the two-point correlation function of \lya \ absorber-galaxy pairs.
These authors have shown that, although the \lya \ absorbers are not
distributed
at random with respect to galaxies, they are clustered around galaxies less
strongly than the galaxies themselves are, in particular over length scales of
1.6 to $16 \h50$~Mpc.
\par
\begin{table*}
\caption{Clustering of \lya \ systems and galaxies}
\label{cluster}
\begin{tabular}{llrllrllll}
\noalign{\smallskip}\hline \\
$N_{\rm a}$ & $z_{\rm a}$ & $|\Delta v_{\rm a-a}|$ & $N_{\rm g}$ & $z_{\rm g}$&
d &
 Size & Mag & $|\Delta v_{\rm g-g}|$ & Nature \\
          &             &    \kms              &           &            & kpc&
  kpc $\times$ kpc  &   &     \kms          &        \\
\noalign{\medskip}\hline \\
&&&&&&TON 153 &&& \cr
 2 & 0.5349 & 2050 & 3 & 0.5324 & 1624.0 & 2300$\times$100  & $-21.3$ & 1450 &
\lya  \\
   & 0.5454 &      &   & 0.5397 &  563.5 & & $-22.3$ &      &          \\
   &        &      &   & 0.5398 &  578.4 & & $-22.0$ &      &          \\
 6 & 0.6606 & 3000 & 4 & 0.6639 & 1471.1 & 1500$\times$300 & $-22.4$ & 1400 &
metal    \\
   & 0.6692 &      &   & 0.668  & 1242.4 & & $-23.2$ &      & and \lya \\
   & 0.6716 &      &   & 0.6715 &   77.5 & & $-22.0$ &      &          \\
   & 0.6736 &      &   & 0.6717 &  415.3 & & $-21.3$ &      &          \\
   & 0.6754 &      &   &        &        & &         &      &          \\
   & 0.6771 &      &   &        &        & &         &      &          \\
 2 & 0.7339 & 1450 &   &        &        & &         &      & \lya     \\
   & 0.7423 &      &   &        &        & &         &      &          \\
 2 & 0.7613 &  920 & 1 & 0.7663 & 2375.7 & & $-21.3$ &      & \lya     \\
   & 0.7667 &      &   &        &        & &         &      &          \\
 2 & 0.8294 & 2750 &   &        &        & &         &      & \lya     \\
   & 0.8463 &      &   &        &        & &         &      &          \\
 2 & 0.9653 & 1900 &   &        &        & &         &      & \lya     \\
   & 0.9778 &      &   &        &        & &         &      &          \\
&&&&&&3C 351 &&& \\
 ? &        &      & 2 & 0.0708 &  296.2 & 220$\times$70 & $-20.0$ & 140 &
O{\sc i} geoc. \\
   &        &      &   & 0.0713 &   62.9 & & $-18.0$ &      &          \\
 1 &        &      & 2 & 0.0860 &  325.3 & 200$\times$160& $-21.0$ & 1650 &
\lya     \\
   & 0.0920 &      &   & 0.0910 &  125.3 & & $-20.2$ &      &          \\
 ? &        &      & 4 & 0.0962 &  498.9 &250$\times$180 & $-22.4$ &  380 &
C\,{\sc ii} Gal. \\
   &        &      &   & 0.0970 &  692.2 & & $-19.1$ &      &          \\
   &        &      &   & 0.0973 &  405.7 & & $-22.3$ &      &          \\
   &        &      &   & 0.0976 &  389.3 & & $-20.5$ &      &          \\
 2 & 0.1611 & 1670 & 2 & 0.1621 &  722.4 &170 $\times$90 & $-22.9$ & 1570 &
\lya     \\
   & 0.1676 &      &   & 0.1682 &  143.8 & & $-19.4$ &      &          \\
 2 & 0.2216 &  320 & 2 & 0.2217 &  710.6 & 1000$\times$70& $-19.5$ & 1050 &
metal     \\
   & 0.2229 &      &   & 0.2260 &  347.0 & & $-19.8$ &      & and \lya  \\
 3 & 0.3621 & 2080 & 2 & 0.3621 &  189.2 & 120$\times$60 & $-21.5$ & 1930 &
\lya     \\
   & 0.3646 &      &   &        &        & &         &      & and metal\\
   & 0.3716 &      &   & 0.3709 &  301.6 & & $-23.7$ &      &            \\
&&&&&&  H 1821+6419 &&& \\
 1 & 0.1806 &      & 2 & 0.1785 &  672.6 &150 $\times$240 & $-21.0$ & 2750 &
\lya     \\
   &        &      &   & 0.1894 &  952.4 & & $-22.2$ &      &            \\
 1 & 0.2972 &      & 25& 0.2844 &   56.6 &3000$\times$1000 & $-23.3$ &
1100$^{\rm a}$ & metal \\
   &        &      &   &   to   &   to   & &  to   &      &            \\
   &        &      &   & 0.3040 & 1534.4 & & $-20.7$ &      &           \\
\noalign{\medskip}\hline
\end{tabular}
\smallskip\noindent \\
$^{\rm a}$ This is the velocity dispersion of the cluster.
\end{table*}
The \lya \ absorption systems show clustering on velocity scales of 1000 to
3000\kms \ (Bahcall et al. 1994), thus larger than observed for galaxies within
groups and sheets of large-scale structures (Ramella et al. 1989) and closer to
velocity spreads found for cluster or superclusters of galaxies. Furthermore,
about half of the extensive metal-rich absorption systems are accompanied by
such
clumps of neighboring \lya \ absorption lines. Our galaxy sample can be used to
investigate the occurrence of galaxy clumps, on similar velocity scales,
associated with one or more \lya \ absorption lines. To conduct this study, we
select \lya \ absorption lines (including metal-rich systems) satisfying the
same criterion limit as used by Bahcall et al. (1993, 1994), $\Wr \ge
0.24$~\AA.
\par
The clumps of \lya \ lines and associated galaxies, with impact parameters up
to
$1.6 \h50$~Mpc, are listed in Table~\ref{cluster}. The largest grouping of \lya
\
lines is found towards the quasar TON~153. There are five \lya -only lines
at $\za \simeq 0.67 \ll z_{\rm e}$, clustered within 1400\kms\ ; including a
neighboring metal-rich system leads to a total velocity span of about 3000\kms.
At least four of these six absorption systems have an associated galaxy.
Towards
the same quasar, there is also a pair of \lya \ absorbers at $\za \simeq 0.54$
with three associated galaxies (of which two form an interacting pair) spanning
about 1500\kms. In addition the sample includes one triplet and six pairs of
\lya
\ lines spanning from 320 to 2950\kms. The triplet and one of the pairs include
one metal-rich system. For about half of these \lya \ absorbers, there is a
detected, associated galaxy.
\par
The sample also contains groups of galaxies associated with a single \lya \
absorption line. There is a rich cluster of galaxies detected towards the
radio-quiet quasar H~1821+6419, with 25 galaxies at $\zg \simeq \za \simeq
z_{\rm
e} = 0.297$ (including the galaxies observed by Schneider et al. 1992),
spanning
about 4000\kms \ and with a velocity dispersion of
1100\kms. The second largest grouping of galaxies is detected towards the
quasar 3C~351, with four galaxies at $\left< \zg \right> = 0.0973 $ and
spanning
only 380\kms, indicative of a physical group. Unfortunately, the associated
\lya
\ absorption is expected at $\lambda _{\rm obs} = 1334.0$~\AA \  and would be
heavily blended with the strong Galactic C\,{\sc ii}($\lambda _{\rm obs} =
1334.5$~\AA)
absorption. The pair of galaxies at $\zg = 0.0710$ is also another inconclusive
case : the impact parameters for both galaxies are $D < 350\h50$~kpc and their
relative velocity is $\Delta v = 140$\kms , but the associated \lya \
absorption
line would be blended with the strong O\,{\sc i} Geocoronal emission line.
The interacting pair of galaxies at $\zg = 0.5398$ is associated to a single
\lya \ absorber and it should be embedded in a  supergiant halo of radii of
order
$600 \h50$~kpc. This could also be the
case for the two galaxies at the same redshift $\zg = 0.6716 = \za$, although
there is a rich clump of neighboring \lya \ absorption lines and the two
galaxies could alternatively be associated with different \lya \ absorbers.
There are two pairs of galaxies with impact parameters smaller than $1
\h50$~Mpc,
but their velocity relative to the absorber is large ($> 750$\kms), and thus
they
do not represent cases of \lya \ absorber-galaxy associations.
\par
\begin{figure}
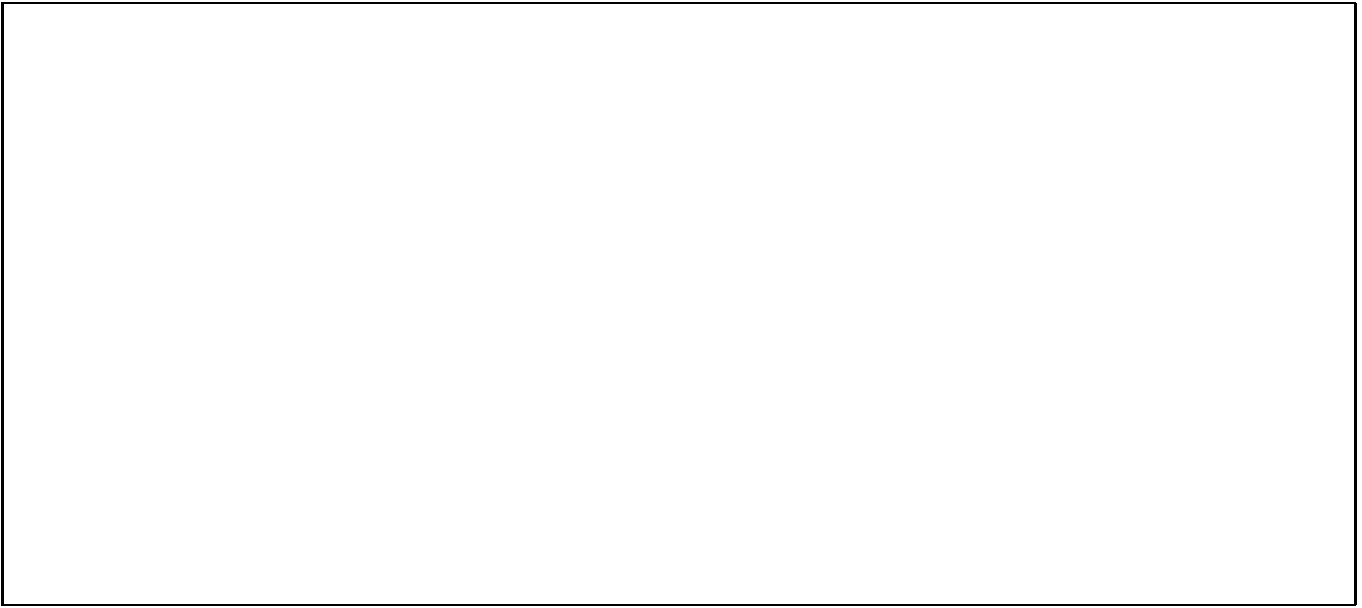

\picplace{8cm}
\caption[]{Galaxy surface-density around TON~153.The full line is the mean
density $\rho_0$, the dotted line is 1 $\sigma$ below $\rho _0$, the
short-dashed
line is 1 $\sigma$ above $\rho _0$ and the dot-dashed line is 2$\sigma$ above
$\rho _0$. The cross
indicates the quasar location}
\label{q13de}
\end{figure}
\begin{figure}
\picplace{8cm}
\caption[]{Same as  Fig.~\ref{q13de} for 3C~351}
\label{q17de}
\end{figure}
\begin{figure}
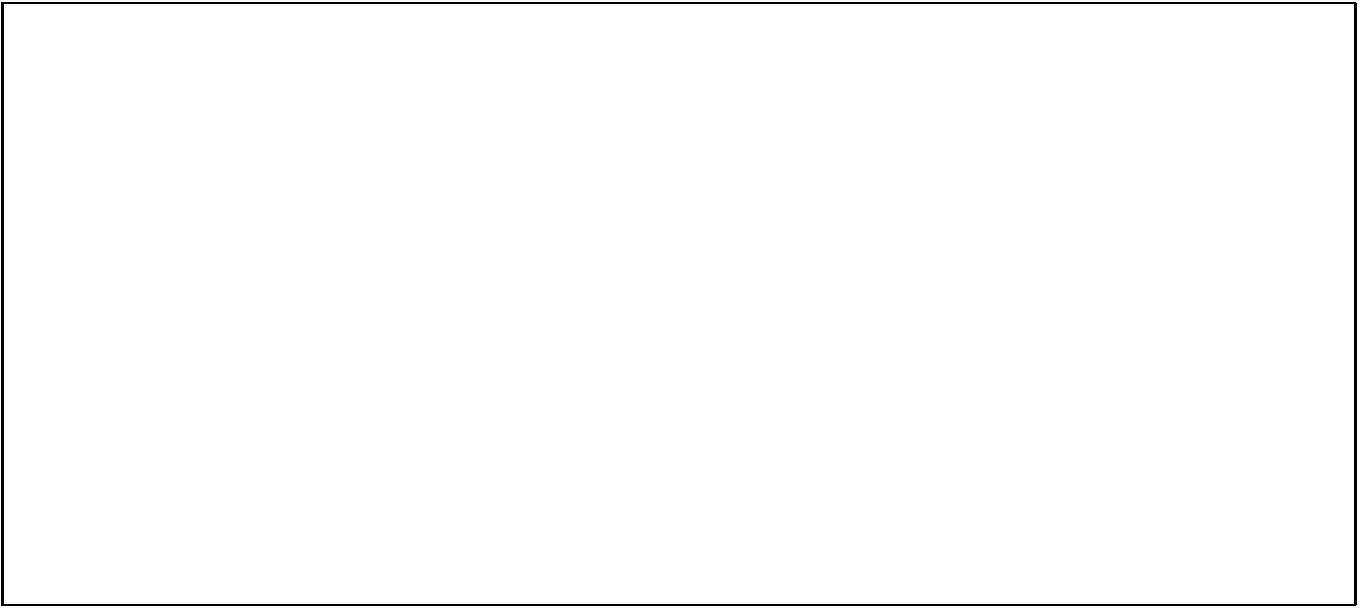

\picplace{8cm}
\caption[]{Same as Fig.~\ref{q13de} for  H~1821+6419,  and long-dashed lines
for
3 to 8 $\sigma$ above $\rho _0$}
\label{q18de}
\end{figure}
To further study the presence of galaxy clustering in the quasar fields, we
construct the galaxy surface-density map for the different fields. At each
point,
this density is calculated by summing the contribution of all the galaxies in
the
field. A two-dimensional Gaussian function is assigned to each galaxy. This
Gaussian function is normalized and centred on the galaxy position. The field
galaxy
surface-density, averaged over the whole sky, to our limiting apparent
magnitude $m_{\rm r, lim} = 22.5$,  $\langle\rho\rangle$, is derived
by integrating the exponential law given by Le Brun et al. (1993) up  to
$m_{\rm r, lim}$. This leads to a characteristic, projected distance between
galaxies
$\langle d \rangle = \langle \rho \rangle ^{-{1 \over 2}}$, which is used as
the
HWHM of the Gaussian function. For each field, we then estimate the average
galaxy surface-density
for $\mr \le m_{\rm r, lim}$, $\rho _0$. In Figs.~\ref{q13de},
\ref{q17de} and \ref{q18de}, the isodensity contour at $\rho _0$ is plotted as
a
solid line. The standard deviation is
\begin{equation}
\sigma = {\sqrt{n} \over s},
\end{equation}
where $n$ is the average number of galaxies in a cell of size $2\langle
d\rangle$, and $s$ is the surface of such a cell.  In the field of H~1821+6419,
these quantities have been derived from an area which excludes the galaxy
cluster, since it would otherwise lead to overestimating  $\rho _0$.  The
contour
levels are plotted as dotted lines for $\rho _0 - 1\sigma$, small-dashed lines
for $\rho _0 + 1\sigma$, dot-dashed lines for  $\rho _0 + 2\sigma$, and for the
contours at $3\sigma$ and above as long-dashed lines. The only strong feature
appearing in these surface-density maps is the cluster at the redshift of the
quasar H~1821+6419, which is a $8\sigma$ overdensity. There is no overdensity,
even  at the $2\sigma$ level, in the field of 3C~351, while there is a group
west to the image of TON~153 detected at the $2\sigma$ level. This overdensity
contains some galaxies at $z \simeq 0.67$. At this  redshift, our magnitude
limit corresponds to L$^{\star}$ galaxies, and deeper imaging over a large
field is needed to confirm this overdensity.
\section{Conclusions}
The present survey of galaxies in the fields around quasars at intermediate
redshift, combined with previous similar studies, leads to a sample of 32  \lya
\
absorber-galaxy associations and 11 galaxies without associated \lya \
absorption.
The redshift coincidences for the associations is statistically significant.
This
survey favors the existence of two populations among the \lya \ absorber-galaxy
associations at $\left< \za \right> \sim 0.32 $ and have direct implications
regarding the size of gaseous galactic halos.
The continuity in impact parameters and H\,{\sc i} column densities between
metal-rich and \lya -only absorber-galaxy associations suggests that the
\lya \ absorbers are part of coherent  structures, and the results of the
search
for various correlations imply  that only a limited fraction of these absorbers
are physically associated with normal galaxies. Almost invisible galaxies with
low
surface brightness and smaller initial central column density of gas than
ordinary
galaxies could be responsible for a large fraction of \lya \ absorbers
(Salpeter \& Hoffman 1995). This class of galaxies could reside in filaments
which
also contain some ordinary galaxies and galaxy groups.
\par
The link with the large-scale structures of the universe is suggested by 1) the
similarity between the narrow distribution of the relative velocity,
$|\Delta v|$, of the \lya \ absorber-galaxy associations (HWHM = 120 \kms)
and the local thickness of  large-scale structures, 2) the lack of
correlation between the impact parameter of the galaxies and their
luminosity or $|\Delta v|$, 3) the \lya \ absorber-galaxy  correlation function
which is positive, although not as strong as the galaxy-galaxy correlation
function on large scales (Morris at al. 1993). The small values of $|\Delta v|$
found recently by Dinshaw et al. (1995) for \lya \ absorbers at intermediate
redshift toward a close quasar-pair are similar to  the HWHM of the relative
velocity distribution of absorber-galaxy associations. As pointed out by these
authors common \lya \ absorptions detected in quasar-pairs do not allow to
discriminate directly between single coherent clouds and correlated but
distinct
features. The dimensions inferred from these observations constrain either
galactic halo sizes or correlation lengths within large-scale structures.
\par
\begin{figure}
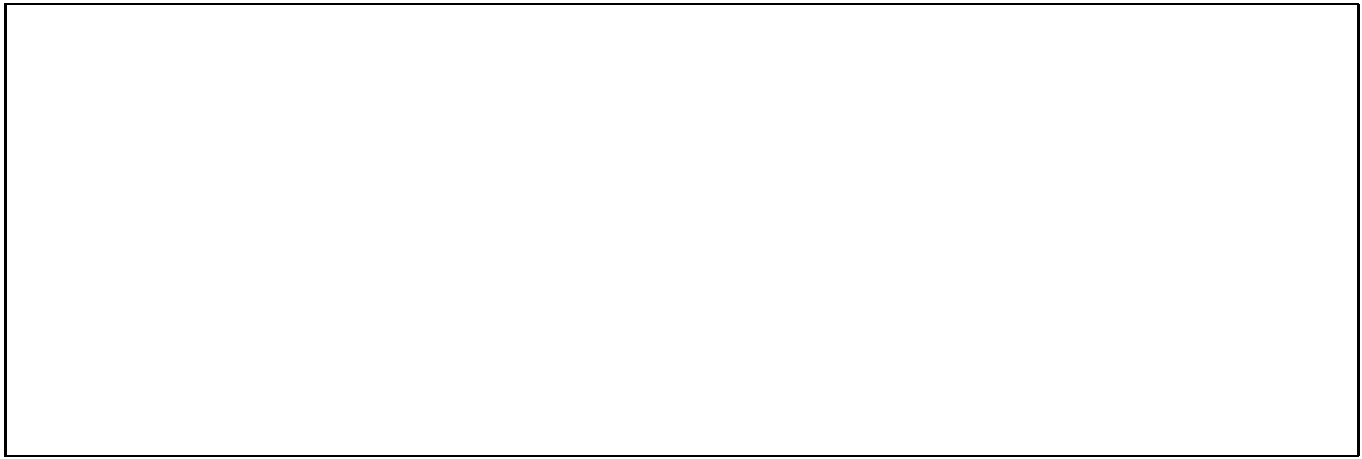

\picplace{6cm}
\caption[]{Cumulated fraction of \lya \ absorber-galaxy associations versus
impact
parameter}
\label{frac}
\end{figure}

The gaseous galactic halos traced by Mg\,{\sc ii} and C\,{\sc iv} absorptions
have radii up to $ \sim$ 75-95 $\h50$~kpc and they may extend further out,
their most outer parts giving rise to only \lya \ absorption. The
overall size of these halos can be derived from the ratio of the cumulated
number of \lya \ absorber-galaxy associations for impact parameters smaller
than $D$ to the total number of galaxies with impact parameters $< D$, shown
in Fig.~\ref{frac}.
This fraction drops from unity at $D < 175 \h50$ kpc down to
about 0.65 for $D > 220 \h50$ kpc and is then roughly constant up to the
largest scales of our combined sample (1.8 $\h50$ Mpc). Our survey has
substantially increased the number of \lya \ absorber-galaxy
associations at intermediate and large impact parameters, and thus this
break does not extend to as large impact parameters as found by LBTW.
Our results then suggest that the characteristic dimensions implied by the
observations of Dinshaw et al. (1995) are more indicative of correlation
lengths than actual halo radii. The size of \lya \ absorption-selected galactic
halos of luminous field galaxies is about three
times larger than the inner parts traced by  Mg\,{\sc ii} absorption.
Although large, this limited halo size as compared to the range of impact
parameters investigated can account for the weak anti-correlation between $D$
and $\Wr$ found for  the sample of \lya \ absorber-galaxy associations
restricted
to $\Wr \geq$ 0.24~\AA.
\par
The presence of groups and clusters of galaxies can be investigated, even
though
our spectroscopic survey is incomplete. Indeed, clustering can be derived from
galaxy density fluctuations within each field (see Sect.~4.4). The cluster of
galaxies associated with the radio-quiet quasar H~1821+6419 is obviously
present
(see Fig.~\ref{q18de}). There is a single absorption system at $\za \simeq
z_{\rm e}$, and its ionization level is high. This absorber could then be
associated to the quasar and its close environment rather than to the cluster
itself. There is also possibly a group of galaxies west of the
quasar TON~153 detected only at the $2 \sigma$ level, which could be a
geometrical superposition of sheets of large-scale structures at $z \simeq
0.660$ and 0.673. Since the average distance between galaxies  within a sheet
of
large-scale structures is $\simeq 6 \h50$~Mpc (see e.g. Ramella et al. 1989,
1992), i.e. at least 1.5 times our field size (for $\zg = \za \le 0.7667$), the
presence of two or more galaxies in one field with relative velocities smaller
than 500 \kms would reveal the presence of at least a loose group. In our three
fields, aside from an interactive galaxy-pair, there is no single \lya \
absorption unambiguously associated with two or more galaxies. There
are one pair and one group of galaxies spanning at most 380\kms, but there is
no
available information on  the associated \lya \ absorption. The presence of
\lya \ absorbers in a rich galaxy environment is thus not likely, as already
discussed by Morris et al. (1993). This question will be more firmly answered
after completion of our survey.

\end{document}